\documentclass{iopjournal}
\usepackage{amsmath}
\usepackage{amssymb}
\usepackage{placeins}
\usepackage{flafter}
\usepackage{booktabs}
\usepackage{siunitx}

\begin{document}

\articletype{Paper} 

\title{Improving Slow-Roll Estimates in Starobinsky Inflation\\ Using an Analytic Hubble Parameter}

\author{Jose Mathew$^1$\orcid{0000-0002-5168-5156}}

\affil{$^1$Department of Physics, The Cochin College, Kochi 682 002, Kerala, India}

\email{josemathew@thecochincollege.edu.in}


\begin{abstract}
Inflationary observables are naturally formulated in terms of the
Hubble-flow hierarchy. Because exact background solutions are often
unavailable, the Hubble-flow parameters are commonly approximated
using potential slow-roll (PSR) quantities constructed from
derivatives of the inflationary potential. We revisit the accuracy of
this replacement for Starobinsky inflation by mapping analytic
Jordan-frame background variables to their Einstein-frame
counterparts and deriving the Hubble-flow parameters directly. The
leading analytic background is an exact solution of a controlled
deformation of the Starobinsky action. Linearizing the pure
Starobinsky equation about this background yields a closed-form
perturbative correction. Using the same first-order spectral
expressions throughout, the perturbatively corrected background
tracks the numerical evolution of the first two Hubble-flow
parameters more accurately than the next-to-leading-order (NLO) PSR
approximation over the common domain in which the NLO expressions
remain physical. Moreover, the NLO estimate of
\(\widetilde{\epsilon}_1\) reaches a maximum value of approximately
\(0.114\), subsequently decreases, and therefore cannot determine the
end of inflation from \(\widetilde{\epsilon}_1=1\). For the Planck
2018 central value \(n_s=0.9649\), the perturbatively corrected
background gives \(\widetilde{N}_*\simeq55.60\), compared with
\(55.72\) for the converged numerical background and \(56.98\) for
the conventional large-\(\widetilde{N}_*\) estimate. For the adopted
ACT DR6 combined-data value \(n_s=0.974\), it gives
\(\widetilde{N}_*\simeq75.43\), compared with the conventional value
\(76.92\). The analytic construction therefore provides a
self-consistent endpoint and a first-order estimate of the e-fold
number that closely reproduces the converged numerical result.
\end{abstract}

\section{Introduction}
Cosmic inflation provides a compelling framework for explaining several fundamental features of the observable Universe, including its large-scale homogeneity, isotropy, and near spatial flatness \cite{Guth:1980zm,Linde:1981mu, Albrecht:1982wi}. In addition to addressing these classical problems of standard cosmology, inflation also provides a mechanism for generating the primordial density perturbations that seed the formation of large-scale structure \cite{Mukhanov:1981xt, Hawking:1982cz, Starobinsky:1982ee}. Related applications of modified-gravity field equations to
relativistic fluid configurations are discussed in
Refs.~\cite{khan2023impact,mardan2020new,noureen2022development}. In most models, inflation is driven by a scalar field slowly evolving along its potential, producing a quasi-de Sitter phase of accelerated expansion. The predictions of inflationary models are typically expressed in terms of observable quantities such as the scalar spectral index $n_s$, the tensor-to-scalar ratio $r$, and the amplitude of scalar perturbations $A_s$, which can be compared with increasingly precise cosmological observations \cite{Planck:2018jri}.

Among the many inflationary scenarios proposed in the literature, the Starobinsky model \cite{Starobinsky:1980te} remains one of the most successful and well-motivated. Originally formulated as a modification of general relativity with an $R + R^2$ action, the model naturally generates an inflationary phase driven by higher-curvature corrections to the gravitational action. When expressed in the Einstein frame through a conformal transformation, the theory becomes dynamically equivalent to a scalar field with a characteristic plateau-like potential \cite{Whitt:1984pd,Maeda:1988ab}. This potential leads naturally to slow-roll inflation and predicts values of the scalar spectral index and tensor-to-scalar ratio that are in excellent agreement with current observational constraints \cite{Planck:2018jri}. Detailed analyses of Starobinsky-type models have further explored the robustness and precision of these predictions under various extensions and corrections \cite{Motohashi:2014tra,Cicoli:2018kdo}. For this reason, Starobinsky inflation is often regarded as a benchmark model in the study of inflationary dynamics.
Recent measurements, including those from the Atacama Cosmology Telescope (ACT), suggest a slightly higher central value of the scalar spectral index compared to Planck, further emphasizing the need for precise theoretical predictions~\cite{AtacamaCosmologyTelescope:2025blo}.

In most analyses of inflation, the dynamics of the inflaton field and the resulting predictions for cosmological observables are described using the slow-roll approximation \cite{book:15483}. Within this framework, two related sets of slow-roll parameters are commonly used. The potential slow-roll parameters are defined in terms of derivatives of the inflaton potential, while the Hubble-flow parameters are defined directly from the evolution of the Hubble parameter \cite{Stewart:1993bc,Liddle:1994dx}. At a fixed order in the slow-roll expansion, the standard expressions for inflationary observables can be written in terms of the Hubble-flow hierarchy. Potential slow-roll formulae are obtained by expressing that hierarchy approximately through derivatives of the potential, which greatly simplifies analytic calculations.

The Hubble-flow parameters can be expressed perturbatively in
terms of potential slow-roll quantities, with conversion relations available beyond leading order \cite{Liddle:1994dx}.
Higher-order corrections to inflationary spectra have also been
studied extensively \cite{Stewart:1993bc,Auclair:2022yxs}.
Here we examine how accurately finite potential slow-roll
truncations reproduce the Hubble-flow parameters obtained directly from the Starobinsky background, and whether these truncations determine the end of inflation self-consistently. We keep the spectral expressions at first order throughout, so that the comparison isolates differences associated with the background and slow-roll approximations.

In this work, we revisit this issue in the context of Starobinsky inflation formulated in the
Einstein frame. Instead of estimating the Hubble-flow parameters through the conventional
potential slow-roll approximation, we determine them directly from an analytic approximation
to the Hubble background. This approach is complementary to recent efforts aimed at refining
inflationary predictions through improved analytic and numerical treatments. In particular,
Drees and Xu~\cite{Drees:2025ngb} evaluate the e-folding integral exactly within leading-order
potential slow roll and also give a useful small-$(1-n_s)$ expansion. We keep these two
approximations distinct below and compare them with the Hubble-background construction.
Higher-order calculations of the primordial spectrum have recently been carried out
up to N3LO for Starobinsky inflation, showing that spectral corrections can be relevant at
the percentage level~\cite{Bianchi:2024qyp}. Starting from an analytic approximation for the Hubble parameter
in the Jordan frame, we map the solution to the Einstein frame and use it to compute the
corresponding Hubble slow-roll parameters. This procedure provides analytic expressions for
the inflationary dynamics that can be compared directly with numerical solutions of the
background equations. We validate the analytic background through a residual analysis, showing
that the normalized deviation from the exact $\Lambda = 0$ Starobinsky equations of motion remains small
throughout the observationally relevant inflationary interval, confirming that the approximation is
controlled rather than merely formal.

The analytic construction developed below is tested against a converged numerical evolution of the Starobinsky background and against several commonly used analytic approximations. We also compare it with the exact leading-order potential slow-roll e-folding integral and the corresponding small-\((1-n_s)\) expansion of Drees and Xu~\cite{Drees:2025ngb}, as well as with the Painlev\'e attractor expansion of the Starobinsky equation~\cite{Toyama:2024ugg}. These complementary comparisons allow the accuracy of the background approximation, the endpoint determination, and the subsequent evaluation of the inflationary observables to be assessed separately.

The analysis also distinguishes the order of the potential slow-roll approximation from an additional large-e-fold truncation. Although these procedures are sometimes treated as equivalent, they need not yield the same estimate of the inflationary e-fold number. We find that the inferred result depends on the accuracy of the adopted background, the self-consistent determination of the end of inflation, and any further large-e-fold expansion applied to the observable relations. The detailed definitions, calibrations, and numerical comparisons are presented in the subsequent sections. Recent studies have similarly emphasized the importance of improving inflationary predictions beyond leading-order slow-roll approximations~\cite{Auclair:2022yxs,martin2014encyclopaedia,martin2014best}.

The remainder of the paper is organized as follows.
Section~\ref{sec:2} describes the reconstruction algorithm,
Section~\ref{sec:staro} applies it to Starobinsky inflation and develops the
residual and numerical comparisons, and Section~\ref{sec:conclusion}
summarizes the results.  Appendix~\ref{sec:altsol} records an alternate
exact solution which gives insight into the structure of \({\rm R+R^2}\) gravity.

Throughout this work, we adopt reduced Planck units where $\hbar=c=1$ and $\kappa^2 = 1/M_{\rm P}^2$, with $M_{\rm P}$ denoting the reduced Planck mass. The metric signature is chosen as $(-,+,+,+)$.

\section{An algorithm for obtaining exact solutions in modified gravity}
\label{sec:2}

In this section, we present a general method~\cite{Mathew:2016anx,Mathew:2017lvh} to obtain exact solutions for the field equations of modified gravity theories. 

Here, the generalized $f(R,\phi)$ action~\cite{DeFelice:2010aj,Hwang:1996xh,capozziello2011beyond} we consider is of the form given by
\begin{equation}
    S=\int{\sqrt{-g}d^4x\left[\frac{1}{2\kappa^2}f\left(R,\phi\right)-\frac{\omega}{2}g^{a b}\nabla_a\phi\nabla_b\phi-V\left(\phi\right)\right]}
    \label{eq:action}
\end{equation}

where $\phi$ is a scalar field, $V(\phi)$ is its potential, $R$ is the
Ricci scalar, and $\omega=\omega(\phi)$ is the kinetic coupling. Varying Eq.~(\ref{eq:action}) with respect to the metric and scalar gives
\cite{Hwang:1996xh}
\begin{subequations}
\begin{equation}
\begin{split}
\frac{F}{\kappa^2}G^a_b&=\omega\left(\phi^{;a}\phi_{;b}-
\frac{1}{2}\delta^a_b\phi^{;c}\phi_{;c}\right)-\frac{1}{2}\delta^a_b\left(R\frac{F}{\kappa^2}-\frac{f}{\kappa^2}+2V\right)\\&+\frac{F^{;a}_{\;\;b}}{\kappa^2}-\delta^a_b\frac{1}{\kappa^2}\Box{F},\hfill\\
 \end{split}
 \label{eq:Feq}
 \end{equation}
 Variation with respect to $\phi$ gives
 \begin{equation}
      0=\Box\phi+\frac{1}{2\omega}\left(\omega_{,\phi}\phi^{;a}\phi_{;a}+\frac{1}{\kappa^2}f_{,\phi}-2V_{,\phi}\right),
      \label{eq:eom}
 \end{equation}
\end{subequations}
where $F\equiv F(R,\phi)=\partial_R f(R,\phi)$. No additional matter sector is included. The scalar kinetic and potential contributions are retained in the general reconstruction equations.

 In this section, we are interested in obtaining the exact solutions for the above set of equations of
motion for a spatially flat Friedmann-Robertson-Walker (FRW) background
\begin{equation}
    ds^2=-dt^2+a^2\left(dx^2+dy^2+dz^2\right)
\end{equation}

where $a=a(t)$ is the scale factor and $t$ is Jordan-frame cosmic time.
For the scalar--tensor reconstruction, the identities
$V_{,\phi}=\dot V/\dot\phi$ and
$f_{,\phi}=(\dot f-F\dot R)/\dot\phi$ apply only on a monotonic branch
with $\dot\phi\neq0$.  We take $\omega=+1$ for a canonical scalar and
$\omega=-1$ for a phantom scalar.  The subsequent Starobinsky application
is instead a pure $f(R)$ theory, contains no independent scalar field, and
does not use the scalar equation below.  The FRW equations are
\begin{subequations}
    \begin{equation}
        \begin{split}
          0&= \frac{1}{2}\omega\dot{\phi}^2+3\frac{\ddot{a}}{a}\frac{F}{\kappa^2}+V-\frac{1}{2\kappa^2}f-3\frac{\dot{F}}{\kappa^2}\frac{\dot{a}}{a}
        \end{split}
        \label{eq:00}
    \end{equation}
       \begin{equation}
        \begin{split}
        0&=\frac{1}{2}\omega\dot{\phi}^2-\frac{\ddot{a}}{a}\frac{F}{\kappa^2}-2\frac{F}{\kappa^2}\frac{\dot{a}^2}{a^2}-V+\frac{1}{2\,\kappa^2}f+\frac{\ddot{F}}{\kappa^2}+2\frac{\dot{F}}{\kappa^2}\frac{\dot{a}}{a}
        \end{split}
        \label{eq:ii}
    \end{equation}
    \begin{equation}
        \begin{split}
            0&=\frac{1}{2\kappa^2}\frac{\dot{f}}{\dot{\phi}}-\omega \ddot{\phi} - 3\omega \dot{\phi}\frac{\dot{a}}{a}-3\frac{F}{\kappa^2}\frac{\dot{a}\ddot{a}}{\dot{\phi}a^2}-3\frac{F}{\kappa^2}\frac{\dddot{a}}{\dot{\phi}a}+6\frac{F}{\kappa^2}\frac{\dot{a}^3}{\dot{\phi}a^3}-\frac{\dot{V}}{\dot{\phi}}
        \end{split}
        \label{eq:phi-frw}
    \end{equation}
    \label{Eq:frweqns}
\end{subequations}

\label{sec:Esolns}
Adding Eqs.~(\ref{eq:00})~and~(\ref{eq:ii}) eliminates $V$ and $f$ and gives
\begin{equation}
    0=\omega \dot{\phi}^2\kappa^2+{\ddot{F}}-\frac{\dot{F}\dot{a}}{a}+\frac{2\ddot{a}F}{a}-\frac{2F\dot{a}^2}{ a^2}
    \label{eq:Main}
\end{equation}

Eq.~(\ref{eq:Main}) is a necessary combination of the FRW equations,
not a complete independent system.  After specifying $a(t)$ and, when present,
$\phi(t)$, it becomes a differential equation for $F(t)$.  A~proposed
reconstruction $F(R,\phi)$ must satisfy $F[R(t),\phi(t)]=F(t)$ along the 
chosen background.  Integrating $F=\partial f/\partial R$ then determines 
$f(R,\phi)$ up to a function of $\phi$, after which one of the original FRW equations determines $V(\phi)$.
The resulting $f(R,\phi)$ and $V(\phi)$ must finally be substituted back into
the remaining independent FRW equation and, on a scalar branch, into
Eq.~(\ref{eq:phi-frw}).  This back-substitution is the required consistency
check of the reconstruction.

\section{Starobinsky Inflation}
\label{sec:staro}
The Starobinsky model is a pure $f(R)$ theory without an independent scalar
field:
\begin{equation}
    S=\int d^4x\sqrt{-g}\;\;\frac{1}{2\,\kappa^2}\;f(R)\,,
\end{equation}
where $f(R)=R+\beta R^2$.

\noindent
\textit{Notation.}
Tilded quantities, such as $\widetilde{\epsilon}_1$,
$\widetilde H$, $\widetilde\phi$, and $\widetilde N_*$, denote
Einstein-frame variables, whereas untilded quantities denote
Jordan-frame variables. The variables $t$ and $\widetilde t$ are
cosmic times in the Jordan and Einstein frames, respectively;
time-translation invariance makes only their additive origins
arbitrary. Here $t_*$ denotes the Jordan-frame time corresponding to
horizon crossing, while $t_f$ denotes the Jordan-frame time at the end
of inflation.

We use
\(\widetilde N(t)\equiv\ln\widetilde a(t)\)
as the forward e-fold time. Hence,
\begin{equation}
\widetilde N_*
=
\widetilde N(t_f)-\widetilde N(t_*)
=
\ln\left(
\frac{\widetilde a_f}{\widetilde a_*}
\right).
\end{equation}
where $\widetilde a_*$ and $\widetilde a_f$ are the Einstein-frame
scale factors evaluated at $t_*$ and $t_f$, respectively. The
inflationary branch evolves from more negative to less negative
Jordan-frame time, with $t_*<t_f<0$.

\subsection{\bf An analytic background solution}
We solve Eq.~(\ref{eq:Main}) for $F$ after removing the independent scalar
field.  Begin with
\begin{equation}
a(t)=a_0\exp\!\left[H_0(qt+C_0)+H_1(qt+C_0)^2\right],
\qquad q=\pm1.
\end{equation}
To display its relation to the simpler form used below, let $y=qt+C_0$.
Completing the square gives
\begin{equation}
H_0y+H_1y^2
=H_1\left(y+\frac{H_0}{2H_1}\right)^2
-\frac{H_0^2}{4H_1}.
\end{equation}
The final constant is absorbed into $a_0$. Choosing the time orientation
$q=+1$, defining $u=t+C_0+H_0/(2H_1)$, and subsequently renaming $u$ as
$t$ yields $a(t)=a_0e^{H_1t^2}$ and $H(t)=2H_1t$.  Solving for $F$
then gives

\begin{equation}
    \begin{split}
        &F(R(t))=\\&{C_1} \left(-2 t \sqrt{\mathit{H_1}}\, {\mathrm e}^{\mathit{H_1} \,t^{2}}+2 \left(\mathit{H_1} \,t^{2}-\frac{1}{2}\right) \mathrm{erfi}\! \left(t \sqrt{\mathit{H_1}}\right)\sqrt{\pi}\right)\\&+{C_2} \left(2 \mathit{H_1} \,t^{2}-1\right)
    \end{split}
    \label{eq:genstaroF}
\end{equation}
where $C_1$ and $C_2$ are integration constants and
$\operatorname{erfi}(x)=-i\operatorname{erf}(ix)
=2\pi^{-1/2}\int_0^x e^{u^2}du$. We now restrict to the polynomial
quadratic branch by setting $C_1=0$; a nonzero $C_1$ generates a
non-polynomial $F(R)$ branch that is not analysed here. Thus
\begin{equation}
F(R(t))=C_2(2H_1t^2-1).
\end{equation}
For this scale factor, the Ricci scalar is
$R=48H_1^2t^2+12H_1$. Hence the reconstructed action is
\begin{equation}
f(R)=R+\beta R^2+\Lambda.
\label{eq:starosol}
\end{equation}

The additive constant $\Lambda$ is determined by the remaining
Friedmann equation, rather than by integrating $F=df/dR$ alone.
For $H=2H_1t$, $F=\frac{2}{3}(1-2H_1t^2)$, and
$\beta=-1/(72H_1)$, the pure-$f(R)$ form of Eq.~(\ref{eq:00}) gives
\[
0
=
3FH^2-\frac{FR-f}{2}+3H\dot F
=
H_1+\frac{\Lambda}{2}.
\]
Thus $\Lambda=-2H_1$. Together with Eq.~(5), this verifies
both independent background equations. The constants are therefore
\begin{align}
C_2=-\frac{2}{3},\\
\beta=\frac{C_2}{48H_1}=-\frac{1}{72H_1}, \\
\Lambda=-2H_1 .
\end{align}

The solution in Eq.~\ref{eq:starosol} is discussed in
Refs.~\cite{Dimitrijevic:2020dzo,Ketov:2022zhp}; here it is recovered by
the reconstruction above.  The form $H=2H_1t$ follows from the explicit
time shift already displayed and simplifies the subsequent mapping.
We emphasize that the constant \(\Lambda\) appearing in Eq.~(\ref{eq:starosol})

is an additive constant in the definition of \(f(R)\), and should not be
identified directly with the conventional cosmological constant. In the
standard Einstein--Hilbert convention one writes
\begin{equation}
f(R)=R+\beta R^{2}-2\Lambda_{\rm phys}.
\end{equation}
The two conventions are therefore related by
\begin{equation}
\Lambda_{\rm phys}=-\frac{\Lambda}{2}.
\end{equation}
For the analytic solution considered here,
\begin{equation}
\Lambda=-2H_{1},
\end{equation}
and hence
\begin{equation}
\Lambda_{\rm phys}=H_{1}.
\end{equation}
Since the inflationary branch requires \(H_{1}<0\), the corresponding
conventional cosmological constant is negative. Throughout this work,
\(\Lambda\) denotes the additive constant in \(f(R)\), while
\(\Lambda_{\rm phys}\) denotes the cosmological constant in the standard
Einstein--Hilbert convention.
For the pure Starobinsky target theory, the usual background viability
conditions are $F>0$ and $f_{,RR}=2\beta>0$. Since
$\beta=-1/(72H_1)$, the second condition requires $H_1<0$.  For the
leading analytic background,
\begin{equation}
F=\frac{2}{3}\left(1-2H_{1}t^{2}\right)>0
\end{equation}
throughout the inflationary interval. On this branch, expansion requires
$H=2H_1t>0$ and therefore selects $t<0$. Other sign choices are not used
in the inflationary analysis.

For completeness, a different exact solution of quadratic gravity
with a cosmological constant, based on the scale factor
\(a(t)=a_0\,\sqrt{-t}\exp(H_1t^2)\), is summarized in
Appendix~\ref{sec:altsol}. This solution is not used in
the inflationary analysis presented below. 
 
\subsection{Approximation to the pure Starobinsky background}

Throughout the phenomenological analysis, the target theory is the pure
Starobinsky model
\begin{equation}
f_{0}(R)=R+\beta R^{2},
\qquad \Lambda=0 .
\end{equation}
The analytic background
\begin{equation}
H_{\rm app}(t)=2H_{1}t
\end{equation}
is not an exact solution of this target theory. Rather, it is an exact
solution of the nearby deformed model
\begin{equation}
f_{\Lambda}(R)=R+\beta R^{2}+\Lambda,
\qquad
\Lambda=-2H_{1}.
\end{equation}
In the following, this exact solution of the deformed model is used solely
as an analytic trial background for the $\Lambda=0$ Starobinsky dynamics.

Along this background,
\begin{equation}
R=48H_{1}^{2}t^{2}+12H_{1},
\end{equation}
and hence
\begin{equation}
R+\beta R^{2}
=
-32H_{1}^{3}t^{4}
+32H_{1}^{2}t^{2}
+10H_{1}.
\end{equation}
The additive term is subdominant over the inflationary interval considered,
but the hierarchy is not uniform:
\begin{equation}
\frac{|\Lambda|}
     {|R+\beta R^{2}|}
\ll 1
\qquad\hbox{during slow roll}.
\end{equation}
For the $H_{\rm app}$ calibration used below, this ratio is
$1.81\times10^{-5}$ at horizon crossing and rises to
$4.79\times10^{-2}$ at the analytic endpoint.  The deformation is therefore
negligible near horizon crossing but reaches the few-percent level near the
end, which motivates the direct residual and numerical checks.
Nevertheless, the two actions are not identified. The accuracy of the
trial background is assessed directly below by substituting it into the
exact $\Lambda=0$ field equation and by comparing its Einstein-frame
Hubble-flow parameters with the numerical $\Lambda=0$ evolution.

\subsubsection{Perturbative refinement of the analytic background}
\label{sec:perturbative}

Before quantifying the accuracy of $H_{\rm app}=2H_1t$, it is useful to note
that this analytic background can be improved. Writing
$H=H_{\rm app}+\delta H(t)$ and linearizing the exact background equation
\begin{equation}
\mathcal{E}[H]\equiv108\beta H^2\dot H+36\beta H\ddot H-18\beta\dot H^2+3H^2=0
\label{eq:exactEOM}
\end{equation}
to first order in $\delta H$ around $H_{\rm app}$, we find that the coefficient
of $\delta H$ itself vanishes identically, and the linearized equation reduces
to a single first-order equation for $\delta\dot H$:
\begin{equation}
-t\,\delta\ddot H+\left(1-6H_1t^2\right)\delta\dot H=-H_1,
\label{eq:linearized}
\end{equation}
sourced by $\mathcal E[H_{\rm app}]=H_1$. This admits the exact solution
\begin{equation}
\delta H(t)=e^{3|H_1|t^2}\left[\frac{\sqrt{3\pi|H_1|}}{6}\,
\mathrm{erf}\!\left(\sqrt{3|H_1|}\,t\right)+B_1\right]+B_2\,.
\label{eq:deltaHexact}
\end{equation}
The constant \(B_{1}\) is fixed by eliminating the exponentially growing mode
as \(t\rightarrow-\infty\), which gives
\begin{equation}B_1=\sqrt{3\pi|H_1|}/6\,.\end{equation}
This condition removes the growing mode but leaves the additive integration
constant $B_2$ undetermined. The resulting large-$|t|$ asymptotic expansion is
\begin{equation}
\delta H(t)\approx B_2-\frac{1}{6t}+\frac{1}{36|H_1|t^3}-\frac{1}{72|H_1|^2t^5}+\cdots,
\label{eq:deltaHasymptotic}
\end{equation}
verified by direct substitution into Eq.~(\ref{eq:linearized}) to leave a
residual of order $\mathcal{O}(\frac{1}{|H_1|^2t^6})$. Retaining successive
inverse-time terms gives the corrected backgrounds defined below.

We denote the successive truncations of the corrected background by

\begin{align}
H_{\rm corr1}(t)&=2H_1t+B_2^{\rm (corr1)}-\frac{1}{6t},\label{eq:Hcorr1}\\
H_{\rm corr2}(t)&=2H_1t+B_2^{\rm (corr2)}-\frac{1}{6t}
-\frac{1}{36H_1t^3}.\label{eq:Hcorr2}
\end{align}

Here $H_{\mathrm{corr1}}$ retains the first inverse-time correction, whereas $H_{\mathrm{corr2}}$ also retains the term of order $t^{-3}$.

Throughout the following analysis, $H_{\rm corr2}$ is adopted as the
principal analytic background. Unless an explicit superscript or
background label is supplied, the numerical values of $H_1$, $t_*$,
$t_f$, $\widetilde N_*$, and $r$ quoted below refer to
$H_{\rm corr2}$. The vertical lines in Figs.~1--3 mark the common
plotting reference times $t_*^{(\mathrm{corr2})}$ and
$t_f^{(\mathrm{corr2})}$; they are included only as visual reference
lines and are not used to replace the scheme-dependent horizon-crossing
times or endpoints.

The values of $t_*$, $t_f$,
$\widetilde N_*$, and $r$ quoted as the principal predictions of this
work are computed self-consistently from $H_{\mathrm{corr2}}$, using
the horizon-crossing and end-of-inflation conditions obtained from the
same background. The leading background
$H_{\mathrm{app}}=2H_1t$ is retained because it yields compact
closed-form expressions and provides a transparent leading-order
benchmark. The Painlev\'e truncations and the potential slow-roll
results are included for comparison.
After removal of the growing mode, the non-constant part of $\delta H$ decays as
$-1/(6t)$, while the exact solution retains the generic additive
integration constant $B_2$. For any solution or truncation labelled by
$i$ whose affine part has this form, the additive mode can be written as
a shift of the time origin:
\begin{equation}
2H_1^{(i)}t+B_2^{(i)}
=
2H_1^{(i)}
\left(
t+\frac{B_2^{(i)}}{2H_1^{(i)}}
\right).
\end{equation}
Thus, $B_2$ represents the generic additive zero mode of the exact
linearized construction rather than an additional physical parameter.
At the affine level it is equivalent to a shift of the time origin.
For a finite asymptotic truncation, however, an exact global time
translation would shift every occurrence of $t$, and the fitted value of
the additive mode can depend slightly on the number of terms retained.
We therefore denote the separately fitted constants
of the one- and two-correction truncations by
$B_2^{(\mathrm{corr1})}$ and $B_2^{(\mathrm{corr2})}$, respectively.
For the principal background $H_{\rm corr2}$,
$B_2^{(\mathrm{corr2})}$ is determined by matching the truncated
background to the numerical solution of the exact $\Lambda=0$
Starobinsky background equation.


Although $H_{\rm corr2}$ reduces the residual of the
pure Starobinsky background equation during most of the slow-roll
regime, it is an asymptotic correction constructed about the
large-$|t|$ inflationary attractor. Its accuracy deteriorates as the
end of inflation is approached, but it remains the preferred truncation
for the principal analytic prediction.
The leading correction agrees with the first subleading term of the known
Painlev\'e attractor expansion of the Starobinsky equation
\cite{Toyama:2024ugg}. The agreement does not, however, persist at
the next order. Written in our parameterization, the Painlev\'e expansion
truncated at the same order reads

\begin{equation}
H_{\mathrm{Pain}}^{(3)}(t)
=
2H_1(t-t_0)
-\frac{1}{6(t-t_0)}
-\frac{4}{108H_1(t-t_0)^3}.
\label{eq:painlevenlo}
\end{equation}
Using the convention that $n$ denotes the number of terms retained in
the Painlev\'e series, the first partial sum has the same functional form as
$H_{\mathrm{app}}$ after shifting the origin of time, while the second is
$H_{\mathrm{Pain}}^{(2)}
=2H_1(t-t_0)-1/[6(t-t_0)]$. At the level of the formal expansions and
for a common value of $H_1$, this second partial sum has the same two
coefficients as $H_{\mathrm{corr1}}$. The separately calibrated curves
need not coincide because their fitted values of $H_1$ and their
time-origin conventions differ.


For each scheme $i$, the numerical solution is initialized at
\begin{equation}
t_{\mathrm{init}}^{(i)}=1.25t_*^{(i)},
\qquad
H_{\mathrm{num}}^{(i)}(t_{\mathrm{init}}^{(i)})
=2H_1^{(i)}t_{\mathrm{init}}^{(i)},
\qquad
\dot H_{\mathrm{num}}^{(i)}(t_{\mathrm{init}}^{(i)})=2H_1^{(i)}.
\label{eq:numerical-initialization}
\end{equation}

Two distinct numerical trajectories are used below. For each analytic
scheme $i$, $H_{\rm num}^{(i)}$ denotes the exact numerical trajectory
constructed with the same scheme-specific value $\beta^{(i)}$; these
trajectories are used for the relative-error comparisons. Separately,
$H_{\rm num}$ without a scheme label denotes the independently calibrated
numerical-background benchmark shown in the left panel of Fig.~1 and
reported in Table~2. Its Einstein-frame calibration and endpoint
prescription are described in Sec.~\ref{sec:hubble-slow-rol}.
The inflationary branch is an attractor. Accordingly, the numerical
integration is initialized sufficiently early using the asymptotic data
specified above. Initial data lying within the corresponding basin of
attraction converge towards this branch before horizon crossing. 
The same asymptotic initialization prescription is applied separately to each scheme using its own values of \(H_1^{(i)}\), \(\beta^{(i)}\), and \(t_*^{(i)}\).

For each analytic background $H_i$ possessing the generic additive
mode $B_2$, we use $N_{\mathrm{fit}}=400$ uniformly spaced points
$t_j^{(i)}\in[0.98t_{\mathrm{init}}^{(i)},t_*^{(i)}]$ and determine
the corresponding realization $B_2^{(i)}$ by minimizing
\begin{equation}
\chi_i^2\!\left(B_2^{(i)}\right)
=
\frac{1}{N_{\mathrm{fit}}}
\sum_{j=1}^{N_{\mathrm{fit}}}
\left[
H_{\mathrm{num}}^{(i)}\!\left(t_j^{(i)}\right)
-
H_i\!\left(t_j^{(i)};B_2^{(i)}\right)
\right]^2.
\label{eq:genericmatching}
\end{equation}
Here $i$ labels the analytic background, and
$H_{\mathrm{num}}^{(i)}$ denotes the numerical trajectory generated
using the same scheme-specific values of $H_1^{(i)}$ and
$\beta^{(i)}=-1/(72H_1^{(i)})$. The symbol $B_2$ denotes the generic
additive zero mode, whereas $B_2^{(i)}$ denotes its value obtained for
the particular background $H_i$.


Since $B_2^{(\mathrm{corr2})}$ enters $H_{\rm corr2}$
additively, its minimizing value is equivalently
\begin{equation}
B_2^{(\mathrm{corr2})}
=
\frac{1}{N_{\rm fit}}
\sum_{j=1}^{N_{\rm fit}}
\left[
H_{\rm num}^{(\mathrm{corr2})}
\left(t_j^{(\mathrm{corr2})}\right)
-2H_1^{(\mathrm{corr2})}t_j^{(\mathrm{corr2})}
+\frac{1}{6t_j^{(\mathrm{corr2})}}
+\frac{1}{
36H_1^{(\mathrm{corr2})}
\left(t_j^{(\mathrm{corr2})}\right)^3}
\right].\label{eq:B2}
\end{equation}
Because the matching interval depends on
$H_1^{(\mathrm{corr2})}$ and $t_*^{(\mathrm{corr2})}$, while the
horizon-crossing solution depends weakly on
$B_2^{(\mathrm{corr2})}$, the matching
condition above and the horizon-crossing conditions
in Eqs.~(\ref{eq:nscondition}) and~(\ref{eq:Ascondition}) are iterated
to a fixed point. The iteration is stopped when
$B_2^{(\mathrm{corr2})}$, $H_1^{(\mathrm{corr2})}$, and
$t_*^{(\mathrm{corr2})}$ are stable beyond the displayed precision.
This procedure gives
\begin{equation}
B_2^{(\mathrm{corr2})}=-6.20707\times10^{-8}M_{\rm P}.
\label{eq:B2corr}
\end{equation}
The unrounded value
$B_2^{(\mathrm{corr2})}=-6.2070737133\times10^{-8}M_{\rm P}$ is retained
in the numerical calculation.
Thus $B_2^{(\mathrm{corr2})}$ fixes the relative time origin between the truncated
analytic background and the numerical trajectory and does not
represent an additional physical parameter.  For the three-term Painlev\'e background
$H_{\rm Pain}^{(3)}$ defined in Eq.~(\ref{eq:painlevenlo}), the
translation parameter $t_0$ is determined by the analogous
least-squares condition
\begin{equation}
t_0
=
\operatorname*{arg\,min}_{\tau}
\left\{
\frac{1}{N_{\rm fit}}
\sum_{j=1}^{N_{\rm fit}}
\left[
H_{\rm num}^{(\mathrm{Pain})}
\left(t_j^{(\mathrm{Pain})}\right)
-
H_{\rm Pain}^{(3)}
\left(t_j^{(\mathrm{Pain})};\tau\right)
\right]^2
\right\}.
\label{eq:t0fit}
\end{equation}
Here $H_{\rm num}^{(\mathrm{Pain})}$ is the numerical trajectory
generated using the Painlev\'e-specific values
$H_1^{(\mathrm{Pain})}$ and
$\beta^{(\mathrm{Pain})}=-1/[72H_1^{(\mathrm{Pain})}]$.
The fit uses the same $N_{\rm fit}=400$ uniformly spaced points,
with
$t_j^{(\mathrm{Pain})}\in
[0.98t_{\rm init}^{(\mathrm{Pain})},t_*^{(\mathrm{Pain})}]$,
as in Eq.~(\ref{eq:genericmatching}). The resulting value of $t_0$ is reported
in Table~\ref{tab:schemes}.

As a robustness check, setting $B_2^{(\mathrm{corr2})}=0$ gives
$\widetilde N_*=55.60028$ and $r=0.00355218$, showing that the quoted
predictions are insensitive to the matching convention.

\begin{figure*}[!htbp]
    \centering
    \includegraphics[width=0.99\textwidth]{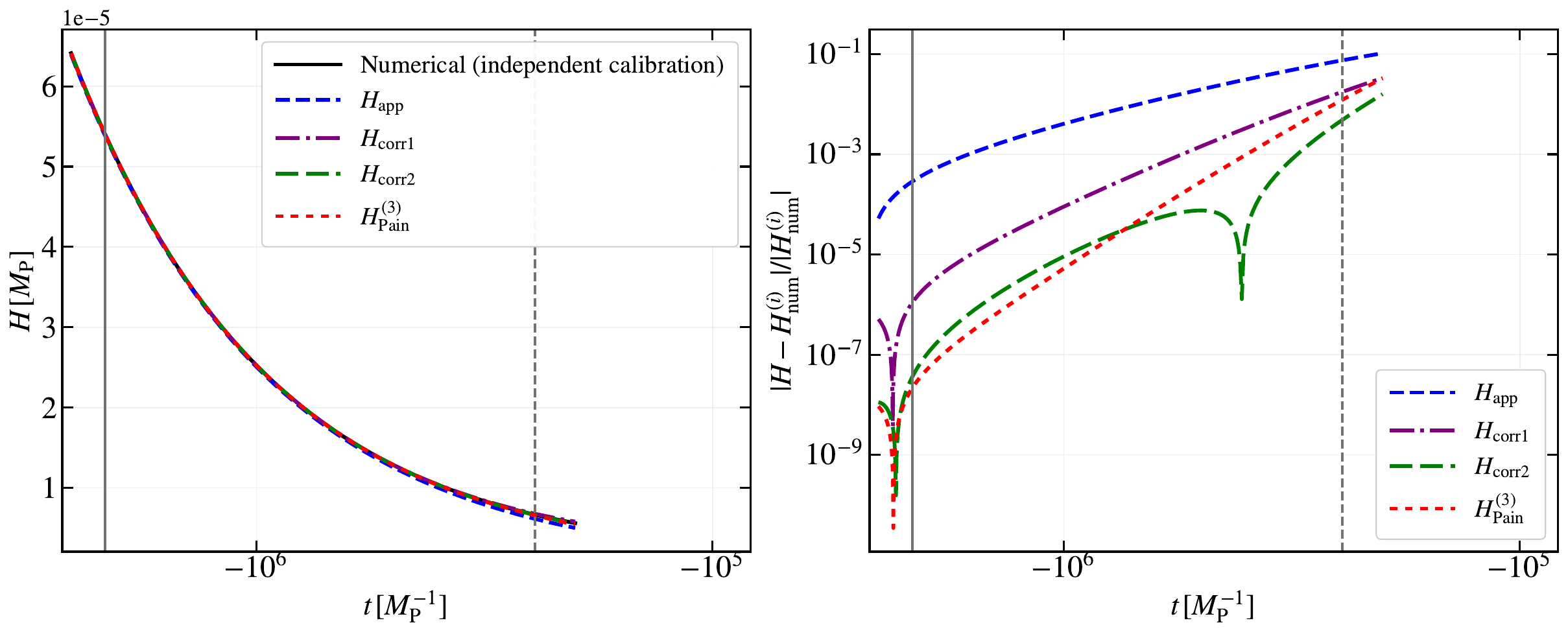}
   \caption{
Comparison of the numerical Jordan-frame Hubble parameter with four
analytic backgrounds. The left panel shows $H(t)$ for the independently
calibrated numerical solution together with $H_{\rm app}$,
$H_{\rm corr1}$, $H_{\rm corr2}$, and $H_{\rm Pain}^{(3)}$.
The right panel shows
$|H_i-H_{\rm num}^{(i)}|/|H_{\rm num}^{(i)}|$, where
$H_{\rm num}^{(i)}$ is the converged numerical trajectory of the exact
background equation, integrated with
the same $\beta^{(i)}=-1/(72H_1^{(i)})$ as analytic scheme $i$.
The corrected truncations are defined in
Eqs.~(\ref{eq:Hcorr1}) and~(\ref{eq:Hcorr2}), and the three-term
Painlev\'e truncation is given in Eq.~(\ref{eq:painlevenlo}).
The grey solid and dashed lines mark, respectively,
$t_*^{\rm corr2}=-2.1480885768\times10^6M_{\rm P}^{-1}$ and
$t_f^{\rm corr2}=-2.4531970802\times10^5M_{\rm P}^{-1}$.
They are reference lines; the predictions in the tables use each
scheme's own calibration and endpoint.
}
    
    \label{fig:Hcomp}
\end{figure*}

The linearized series and the Painlev\'e attractor truncation are compared
with the numerical background in Fig.~\ref{fig:Hcomp}.

\subsubsection{Residual analysis of the analytic background approximation}
\label{sec:residual}
The leading analytic background $H_{\rm app}$ is obtained from the
solution of the slightly deformed action
\begin{equation}
f(R)=R+\beta R^2+\Lambda,
\end{equation}
with \(\Lambda=-2H_1\). Since the inflationary observables discussed in this
work are ultimately interpreted within the pure Starobinsky model,
\begin{equation}
f(R)=R+\beta R^2,
\end{equation}
it is important to quantify how accurately the analytic background satisfies
the exact \(\Lambda=0\) equations of motion. For this purpose, we perform a
residual analysis.

Let the exact background equation for pure Starobinsky gravity be written
schematically as
\begin{equation}
\mathcal{E}[H]=0 .
\end{equation}
For any analytic trial background $H_i(t)$, we define its residual as
\begin{equation}
\mathcal{R}_i(t)=\mathcal{E}[H_i(t)].
\end{equation}
For the leading background, this notation gives
$\mathcal R_{\rm app}(t)=\mathcal E[H_{\rm app}(t)]$.

If \(H_{\rm app}(t)\) were an exact solution of the pure Starobinsky equations,
then \(\mathcal{R}_{\rm app}(t)\) would vanish identically. Since the solution is instead
obtained from the \(\Lambda\neq 0\) deformation, the residual provides a direct
measure of the error introduced by using this analytic background as an
approximation to the \(\Lambda=0\) dynamics.

In order to obtain a dimensionless measure of the accuracy, we normalize the
residual by the sum of the absolute magnitudes of the individual terms entering
the background equation. Thus we define
\begin{equation}
\Delta_{{\rm res},i}(t)
=
\frac{|\mathcal{R}_i(t)|}
{\sum_a |\mathcal{E}_a[H_i(t)]|},
\end{equation}
where \(\mathcal{E}_a\) denotes the separate terms contributing to
\(\mathcal{E}\). This normalization avoids artificial cancellations among
different terms and gives a conservative estimate of the fractional violation
of the exact \(\Lambda=0\) equation.
As a normalization check, we also define
\begin{equation}
\widehat\Delta_{{\rm res},i}(t)
=\frac{|\mathcal E[H_i(t)]|}{3H_i^2(t)}.
\end{equation}

Neither normalization provides, by itself, an upper bound on the
errors in the background solution, its derivatives, or the
inflationary observables. The normalized residual measures only
the violation of the background equation relative to the chosen
normalization. The accuracy of the analytic backgrounds and their
associated Hubble-flow parameters is therefore assessed separately
through direct comparisons with numerical solutions of the pure
Starobinsky theory, as shown in Figs.~\ref{fig:Hcomp} and~\ref{fig:HvsV} and Table~\ref{tab:schemes}. For $H_{\rm app}$, the alternative residual measure evaluated at the
common reference times $t_*^{(\mathrm{corr2})}$ and
$t_f^{(\mathrm{corr2})}$ is
$1.44\times10^{-3}$ and $1.10\times10^{-1}$, respectively. The term-by-term
normalization and this alternative measure give the same qualitative
ordering, while also showing that the approximation becomes less accurate
near the end of inflation.

\begin{figure*}[!htbp]
    \centering
    \includegraphics[width=0.99\textwidth]{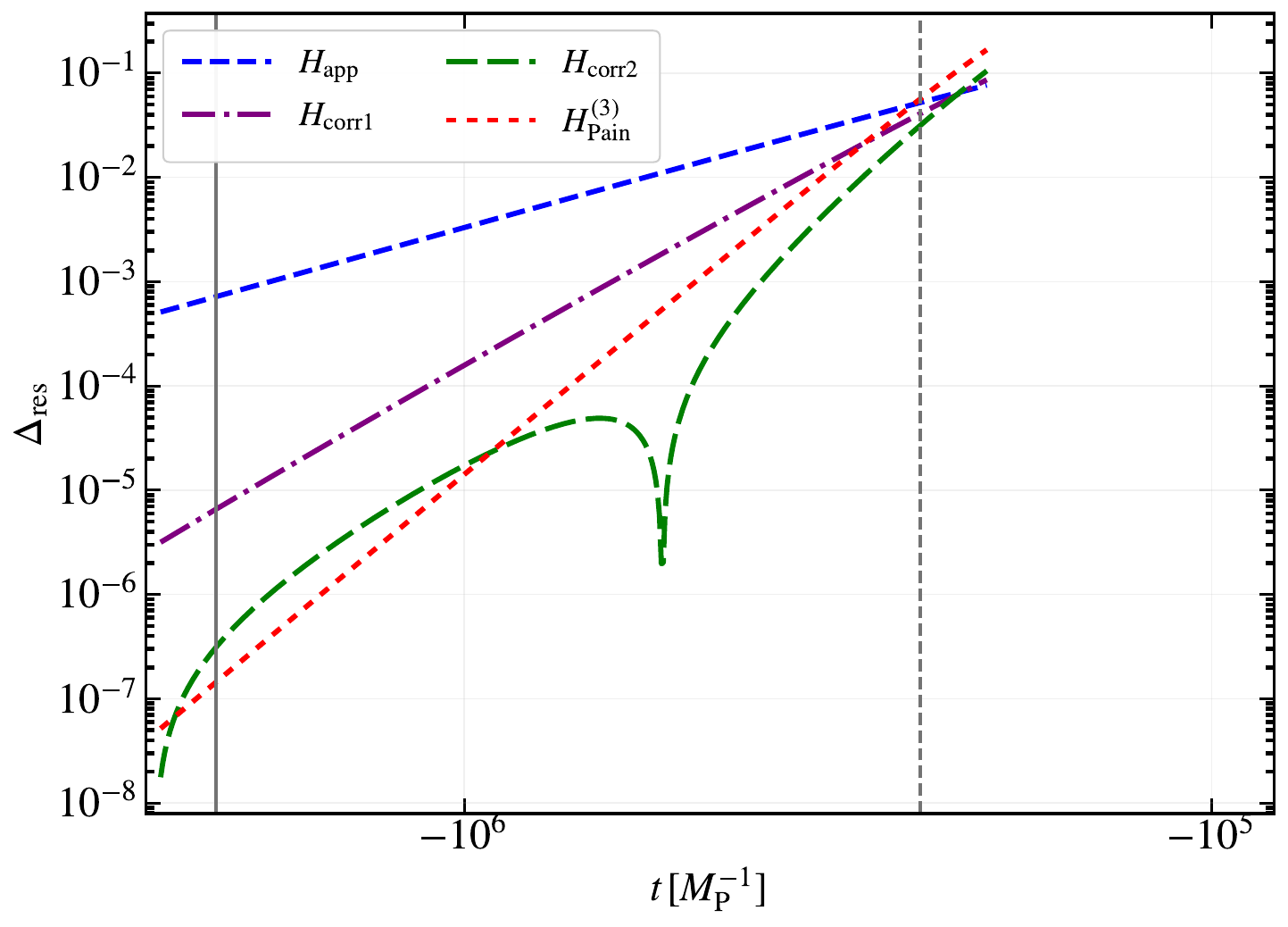}
  \caption{
Term-normalized residual $\Delta_{{\rm res},i}$ obtained by substituting
$H_{\rm app}$, $H_{\rm corr1}$, $H_{\rm corr2}$, and
$H_{\rm Pain}^{(3)}$ into the pure $\Lambda=0$ Starobinsky equation.
Each curve uses that scheme's own $H_1$ and
$\beta=-1/(72H_1)$. At the common reference time
$t_*^{\rm corr2}$, the residuals are, respectively,
$7.19849\times10^{-4}$, $6.59538\times10^{-6}$,
$3.13964\times10^{-7}$, and $1.44953\times10^{-7}$.
The grey lines mark $t_*^{\rm corr2}$ and $t_f^{\rm corr2}$;
they do not replace the scheme-dependent endpoints used in
Table~\ref{tab:schemes}.
}
    \label{fig:residual}
\end{figure*}

For the pure Starobinsky background equation
\begin{equation}
0=
108\beta H^2\dot H
+36\beta H\ddot H
-18\beta \dot H^2
+3\,H^2 ,
\end{equation}
the corresponding residual is obtained by substituting
\begin{equation}
H=2H_1t,\qquad \dot H=2H_1,\qquad \ddot H=0 .
\end{equation}
The resulting normalized residual is plotted in Fig.~\ref{fig:residual}.
It is small during the observationally relevant slow-roll regime and grows as
the endpoint is approached. Thus the analytic background is a controlled
asymptotic approximation to the pure Starobinsky dynamics, rather than an
exact solution of that target theory.

For the perturbatively corrected background $H_{\rm corr1}=2H_1t-1/(6t)$
introduced in Section~\ref{sec:perturbative}, the same substitution into the
exact background equation gives a residual
\begin{equation}
\mathcal R_{\rm corr1}(t)=\frac{24H_1t^2-1}{36H_1t^4},
\label{eq:Rcorr}
\end{equation}
in contrast to the constant residual
$\mathcal R_{\rm app}(t)=H_1$ obtained for $H_{\rm app}$.
Eq.~(\ref{eq:Rcorr}) applies to the unshifted \(B_2^{(\mathrm{corr1})}=0\) one-correction truncation and is quoted to display its asymptotic scaling. The normalized
curves in Fig.~\ref{fig:residual} are instead evaluated directly for the
fitted backgrounds defined in Eqs.~(\ref{eq:Hcorr1}) and~(\ref{eq:Hcorr2}).
Unlike $\mathcal R_{\rm app}$, both $\mathcal R_{\rm corr1}$ and
$\mathcal R_{\rm corr2}$ decay as $|t|\rightarrow\infty$, which is
the underlying reason the corrected background's advantage over $H_{\rm app}$
is largest at early times and narrows as the end of inflation is approached.
For the fitted curves shown in Fig.~\ref{fig:residual}, the residuals at the
common reference crossing time are $7.19849\times10^{-4}$ for $H_{\rm app}$,
$6.59538\times10^{-6}$ for $H_{\rm corr1}$,
$3.13964\times10^{-7}$ for $H_{\rm corr2}$, and
 $1.44953\times10^{-7}$ for the three-term Painlev\'e truncation. The
corrections therefore reduce the violation of the $\Lambda=0$ equation
substantially during slow roll, although the asymptotic hierarchy deteriorates
near the endpoint.
This residual analysis strengthens the use of the analytic solution in the
subsequent calculation of the Hubble slow-roll parameters. In particular, the
smallness of \(\Delta_{{\rm res},i}\) shows that the violation associated
with using the deformed leading background remains controlled at the level
of the pure $\Lambda=0$ background equation. Therefore, the improved
agreement of the analytic Hubble slow-roll parameters with the numerical
\(\Lambda=0\) evolution, shown below, can be interpreted as a controlled
approximation to the Starobinsky inflationary background rather than as the
dynamics of a distinct model.

The purpose of introducing the small non-zero $\Lambda$ term is therefore to obtain a simple analytic background solution that provides a more accurate approximation to the numerical Starobinsky evolution and its associated Hubble slow-roll parameters. Note that $H = 2H_1t$ is an exact solution of the deformed ($\Lambda=-2H_1$)
model and serves only as the trial background for the $\Lambda=0$ dynamics.
The Hubble slow-roll parameters derived from this trial background are compared
below with the numerical results for the pure Starobinsky action ($\Lambda=0$)
and with their potential slow-roll counterparts; the trial-background results
lie closer to the numerical evolution than the leading-order potential
slow-roll expressions (see~Fig.~\ref{fig:HvsV}). 
\FloatBarrier

\FloatBarrier

\subsection{Mapping the Jordan-frame background to the Einstein frame}
\label{sec:Ef}
In the Einstein frame, the action takes the form~\cite{DeFelice:2010aj}
\begin{equation}
    S=\int{d^4x\sqrt{-\widetilde{g}}\left[\frac{1}{2\kappa^2}\widetilde{R}-\frac{1}{2}\partial_a\widetilde{\phi}\partial^a\widetilde{\phi}-V(\widetilde{\phi})\right]}
\end{equation}
where a tilde denotes an Einstein-frame quantity. The frame map is
\begin{equation}
\begin{split}
&\widetilde a=\sqrt{F}a;\quad d\widetilde t=\sqrt{F}\,dt;\quad \\
&\widetilde\phi=M_{\rm P}\sqrt{\frac{3}{2}}\,\ln F; \quad
\widetilde H=\frac{1}{\sqrt{F}}\left(H+\frac{\dot F}{2F}\right)
\end{split}
\label{eq:rel2frames}
\end{equation}
In the Einstein frame, the Starobinsky model is a single scalar field model with a plateau potential. 

At first order in the slow-roll expansion, the scalar and tensor power spectra for a canonical
single-field model are expressed as follows.

\begin{equation}
\mathcal P_{\mathcal R}(k)
=
A_{s*}\left(\frac{k}{k_*}\right)^{n_s-1},
\qquad
A_{s*}\simeq
\frac{{\widetilde H_*}^4}
{4\pi^2{\dot{\widetilde\phi}_*}^2}.
\label{eq:As-fundamental}
\end{equation}

\begin{equation}
\mathcal{P}_T(k) = A_{t*}\left(\frac{k}{k_*}\right)^{n_t}, \qquad
A_{t*} \simeq \frac{2}{\pi^2}\frac{{\widetilde H_*}^{\,2}}{M_{\rm P}^{\,2}}.
\label{eq:At-fundamental}
\end{equation}
where $\mathcal P_{\mathcal R}$ and $\mathcal P_T$ denote the scalar
and tensor power spectra, respectively; $n_s$ and $n_t$ are the
corresponding scalar and tensor spectral indices; and the subscript
$*$ denotes evaluation at horizon crossing. Overdots on tilded
quantities denote derivatives with respect to the Einstein-frame
cosmic time $\widetilde t$, whereas overdots on untilded quantities
denote derivatives with respect to the Jordan-frame cosmic time $t$.
Throughout this work, the analytic backgrounds and the numerical-background
benchmark are compared using these same first-order spectral expressions.
Accordingly, ``numerical benchmark'' below refers to a converged background
integration evaluated at first spectral order; it does not denote an exact
calculation of the perturbation spectrum. This fixed-order convention isolates
differences caused by the background and slow-roll approximations.

Potential slow-roll parameters are widely used because closed analytic
backgrounds are often unavailable. Here the Jordan-frame background is mapped
directly to the Einstein frame, and the Einstein-frame variables are expressed
parametrically in terms of the Jordan-frame cosmic time $t$ and the model
constant $H_1$. The additive origin of $t$ is conventional, but $t$ itself is
the physical Jordan-frame time.

\subsection{\bf Hubble slow-roll parameters and comparison with potential slow-roll parameters}
\label{sec:hubble-slow-rol}
The potential slow-roll procedure approximates the Hubble-flow hierarchy in
terms of derivatives of the Einstein-frame potential. Our procedure instead
maps an analytic Jordan-frame Hubble background to the Einstein frame and
differentiates the mapped Hubble parameter directly. The residual analysis
above explains the improvement of the corrected background: the inverse-time
terms reduce successive asymptotic contributions to the violation of the
exact $\Lambda=0$ equation, so the resulting background and its derivatives
track the numerical attractor more closely.

We first present the construction used for the principal
$H_{\mathrm{corr2}}$ predictions. The compact closed-form expressions for
$H_{\mathrm{app}}=2H_1t$ are given subsequently as a leading benchmark.
Unless explicitly labelled otherwise, the principal analytic results below
refer to $H_{\mathrm{corr2}}$. Substituting
\begin{equation}
H_{\mathrm{corr2}}(t)
=
2H_1t+B_2^{(\mathrm{corr2})}-\frac{1}{6t}
-\frac{1}{36H_1t^3}
\end{equation}
into the Jordan-frame curvature gives
\begin{align}
\dot H_{\mathrm{corr2}}(t)
&=
2H_1+\frac{1}{6t^2}
+\frac{1}{12H_1t^4}, \\
R_{\mathrm{corr2}}(t)
&=
6\left[
2H_{\mathrm{corr2}}^2(t)
+\dot H_{\mathrm{corr2}}(t)
\right], \\
F_{\mathrm{corr2}}(t)
&=
1+2\beta R_{\mathrm{corr2}}(t),
\qquad
\beta=-\frac{1}{72H_1}.
\end{align}
The corresponding Einstein-frame time and Hubble parameter are
therefore
\begin{align}
\frac{d\widetilde t}{dt}
&=
\sqrt{F_{\mathrm{corr2}}(t)}, \\
\widetilde H_{\mathrm{corr2}}(t)
&=
\frac{1}{\sqrt{F_{\mathrm{corr2}}(t)}}
\left[
H_{\mathrm{corr2}}(t)
+\frac{\dot F_{\mathrm{corr2}}(t)}
{2F_{\mathrm{corr2}}(t)}
\right].
\end{align}
The first two Einstein-frame Hubble-flow parameters obtained from this
same corrected background are
\begin{align}
\widetilde\epsilon_{1,\mathrm{corr2}}(t)
&=
-\frac{1}{
\sqrt{F_{\mathrm{corr2}}}\,
\widetilde H_{\mathrm{corr2}}^2}
\frac{d\widetilde H_{\mathrm{corr2}}}{dt}, \\
\widetilde\epsilon_{2,\mathrm{corr2}}(t)
&=
\frac{1}{
\sqrt{F_{\mathrm{corr2}}}\,
\widetilde H_{\mathrm{corr2}}\,
\widetilde\epsilon_{1,\mathrm{corr2}}}
\frac{d\widetilde\epsilon_{1,\mathrm{corr2}}}{dt}.
\end{align}
At first order in the Einstein-frame Hubble-flow expansion, the
quantities evaluated at horizon crossing are
\begin{align}
n_s
&\simeq
1-2\widetilde\epsilon_{1,\mathrm{corr2}}(t_*)
-\widetilde\epsilon_{2,\mathrm{corr2}}(t_*),
\label{eq:ns-corr2}\\
A_s
&\simeq
\frac{
\widetilde H_{\mathrm{corr2}}^2(t_*)
}{
8\pi^2M_{\rm P}^2
\widetilde\epsilon_{1,\mathrm{corr2}}(t_*)
},
\label{eq:As-corr2}\\
r
&\simeq
16\widetilde\epsilon_{1,\mathrm{corr2}}(t_*).
\label{eq:r-corr2}
\end{align}
For an exact canonical Einstein-frame background,
Eq.~(\ref{eq:As-fundamental}) is equivalent to
Eq.~(\ref{eq:As-corr2}), and the tensor-to-scalar ratio reduces to
Eq.~(\ref{eq:r-corr2}), since the background equations imply
\[
\dot{\widetilde\phi}^{\,2}
=2M_{\rm P}^2\widetilde H^2\widetilde\epsilon_1.
\]
The background $H_{\mathrm{corr2}}$, however, is a truncated analytic
approximation to the pure Starobinsky dynamics and therefore does not
satisfy this identity exactly. To avoid mixing two slightly different
prescriptions, Eqs.~(\ref{eq:ns-corr2})~and~(\ref{eq:As-corr2}) are used
consistently both in fixing $H_1$ and $t_*$ and in reporting the observables.
At horizon crossing,
the difference from evaluating the corresponding scalar-field
expressions directly is only of relative order $10^{-6}$.
Thus all the Einstein-frame quantities entering the principal
predictions are constructed from $H_{\mathrm{corr2}}$; the expressions
for $F$, $\widetilde H$, and the Hubble-flow parameters obtained from
$H_{\mathrm{app}}$ are not used in this calculation.

For the calibration below, we adopt the Planck 2018 central value
\(n_s=0.9649\) and use
\(A_{s*}=10^{-10}e^{3.043}\) as our adopted scalar-amplitude
benchmark. See~Ref.~\cite{Planck:2018jri} for details. The corresponding horizon-crossing conditions
\begin{align}
n_s\!\left(
H_1^{(\mathrm{corr2})},
t_*^{(\mathrm{corr2})};
B_2^{(\mathrm{corr2})}
\right) &= 0.9649,
\label{eq:nscondition} \\
A_s\!\left(
H_1^{(\mathrm{corr2})},
t_*^{(\mathrm{corr2})};
B_2^{(\mathrm{corr2})}
\right) &= 10^{-10}e^{3.043}
\label{eq:Ascondition}
\end{align}
are solved simultaneously using the iteratively matched value of
$B_2^{(\mathrm{corr2})}$. On the expanding inflationary branch $H_{\rm corr2}$,
defined by
\begin{equation}
H_1^{(\mathrm{corr2})}<0,\qquad
t_*^{(\mathrm{corr2})}<0,\qquad
F_{\mathrm{corr2}}(t_*)>0,\qquad
\widetilde H_{\mathrm{corr2}}(t_*)>0,
\end{equation}
the relevant solution is
\begin{align}
H_1^{(\mathrm{corr2})}
&=
-1.2570866004\times10^{-11}M_{\rm P}^2, \\
t_*^{(\mathrm{corr2})}
&=
-2.1480885768\times10^6M_{\rm P}^{-1}.
\end{align}
At horizon crossing, this solution gives
\begin{align}
\widetilde\epsilon_{1,\mathrm{corr2}}(t_*)
&=
2.22008944\times10^{-4}, \\
\widetilde\epsilon_{2,\mathrm{corr2}}(t_*)
&=
0.0346559821,
\end{align}
and hence
\begin{equation}
n_s=0.9649,
\qquad
r=0.00355214310.
\end{equation}

The end of inflation is determined independently from the same
corrected background through
\begin{equation}
\widetilde\epsilon_{1,\mathrm{corr2}}(t_f)=1,
\end{equation}
which can have more than one algebraic root for a finite truncation. We define
$t_f$ as the first root encountered while evolving forward from $t_*$ on the
expanding branch with $F>0$ and $\widetilde H>0$; later roots lie outside the
inflationary interval and are discarded. This prescription gives
\begin{equation}
t_f
=
-2.4531970802\times10^5M_{\rm P}^{-1}.
\end{equation}
The corresponding Einstein-frame number of e-folds is
\begin{align}
\widetilde N_*^{\mathrm{corr2}}
&=
\int_{t_*}^{t_f}
\widetilde H_{\mathrm{corr2}}(t)
\sqrt{F_{\mathrm{corr2}}(t)}\,dt \\
&=
\int_{t_*}^{t_f}
\left[
H_{\mathrm{corr2}}(t)
+\frac{\dot F_{\mathrm{corr2}}(t)}
{2F_{\mathrm{corr2}}(t)}
\right]dt \\
&=
H_1\left(t_f^2-t_*^2\right)
+B_2^{(\mathrm{corr2})}\left(t_f-t_*\right)
-\frac{1}{6}
\ln\left|\frac{t_f}{t_*}\right| \nonumber\\
&\quad
+\frac{1}{72H_1}
\left(
\frac{1}{t_f^2}-\frac{1}{t_*^2}
\right)
+\frac{1}{2}
\ln\left[
\frac{F_{\mathrm{corr2}}(t_f)}
{F_{\mathrm{corr2}}(t_*)}
\right] \\
&=
55.60240507.
\end{align}
For this solution,
\begin{equation}
F_{\mathrm{corr2}}(t_f)=1.84659514,
\end{equation}
which is the value quoted in Table~\ref{tab:schemes}.

The sequence is asymptotic rather than convergent, and endpoint
completeness is not monotonic in the number of terms retained. The
endpoint must therefore be determined directly by solving
\(\widetilde\epsilon_1(t_f)=1\), subject to \(F(t_f)>0\) and
\(\widetilde H(t_f)>0\).

For \(H_{\rm corr1}\), the retained \(t^{-1}\) correction becomes
increasingly important as the endpoint is approached, while the
next \(t^{-3}\) contribution is absent. This produces a partial
cancellation among the terms entering
\(d\widetilde H/d\widetilde t\). Consequently,
\(-d\widetilde H/d\widetilde t\) decreases relative to
\(\widetilde H^2\), and
\(\widetilde\epsilon_{1,\rm corr1}\) turns over after reaching its
maximum:
\begin{equation}
\max\widetilde\epsilon_{1,\rm corr1}
\simeq 0.50101 < 1 .
\label{eq:corr1-no-endpoint}
\end{equation}
Thus, the one-correction truncation cannot determine the end of
inflation self-consistently. When the \(t^{-3}\) term is retained in
\(H_{\rm corr2}\), it changes this balance and prevents the turnover:
\(\widetilde\epsilon_{1,\rm corr2}\) continues to increase and reaches
unity while \(F>0\) and \(\widetilde H_{\rm corr2}>0\). The same direct
endpoint test gives valid physical roots for \(H_{\rm app}\) and the
three-term Painlev\'e background. The failure of \(H_{\rm corr1}\)
therefore reflects the loss of endpoint accuracy in the
one-correction asymptotic truncation, rather than the absence of an
inflationary endpoint in the exact numerical background. Retaining further inverse-time terms from the same first-order perturbative correction may improve the approximation within its
asymptotic regime, but it does not guarantee a uniformly more accurate
endpoint. A systematic improvement near the end of inflation would
require higher-order perturbative corrections or direct numerical
evolution.

For the leading benchmark $H_{\rm app}$, applying
Eq.~(\ref{eq:rel2frames}) and the chain rule
$d\Xi/d\widetilde t=(d\Xi/dt)(dt/d\widetilde t)$ gives
\begin{equation}
\widetilde{H}=\frac{12 H_1^{2}\, t^{3}}
{\sqrt{6 - 12 H_1 t^{2}}\, \left(2 H_1 t^{2} - 1\right)}\label{eq:Hfull}
\end{equation}

\begin{equation}
\widetilde{\phi}=\frac{\sqrt{6}}{2\kappa}\,
\ln\!\left(\frac{2 - 4 H_1 t^{2}}{3}\right)
\end{equation}
\begin{equation}
\frac{d\widetilde{t}}{dt}=\frac{\sqrt{6\left(1 - 2 H_1 t^{2}\right)}}{3}
\end{equation}
\begin{equation}
\widetilde{\epsilon_1}=-\frac{d\widetilde{H}}{d\,t}\frac{d\,t}{d\widetilde{t}}\frac{1}{\widetilde{H}^2}=\frac{3}{4\,H_1^2\,t^4}\label{eq:epsilon1}
\end{equation}
\begin{equation}
\widetilde{\epsilon_2}=\frac{d\widetilde{\epsilon_1}}{d\widetilde{t}}\frac{1}{\widetilde{\epsilon_1} \widetilde{H}}=
-\frac{2 H_1 t^{2} - 1}{H_1^{2} \, t^{4}}
\label{eq:epsilon2}
\end{equation}
\begin{equation}
\widetilde{N}_*
=
\ln\left(\frac{\widetilde{a}_f}{\widetilde{a}_*}\right),
\qquad
\widetilde{N}_*^{(\mathrm{app})}
=
H_1\left(t_f^2-t_*^2\right)
+\frac{1}{2}\ln\left(\frac{F_f}{F_*}\right).
\end{equation}
Using the first-order expressions of
Refs.~\cite{DeFelice:2010aj,baumann2022cosmology}, we obtain
\begin{eqnarray}
k_*=\widetilde a_*\widetilde H_*
=\left[\frac{1}{\sqrt F}\frac{d(\sqrt F\,a)}{dt}\right]_*
=\left(\frac{4 {a_0} \,{e}^{{H_1} \,t^{2}} {H_1}^{2} t^{3}}{2 {H_1} \,t^{2}-1}\right)_*\label{eq:kstar}\\
{A}_{s*}=\left(-\frac{4 t^{10} \mathit{H_1}^{6}}{\pi^{2} \left(2 \mathit{H_1} \,t^{2}-1\right)^{3}\,{M_{\rm P}}^2}\right)_* 
\label{eq:PSinpms}
\end{eqnarray}
Similarly,
\begin{equation}
     {A}_{t*}=\left(-\frac{48 \mathit{H_1}^{4} t^{6}}{\mathit{{M_{\rm P}}}^{2} \left(2 \mathit{H_1} \,t^{2}-1\right)^{3} \pi^{2}}\right)_*\label{eq:PStinpms}
\end{equation}
and the tensor-to-scalar ratio is
\begin{equation}
    {r} = \left(\frac{12}{\mathit{H_1}^{2} t^{4}} \right)_*\label{eq:r}
\end{equation}
The subscript $*$ means that the enclosed quantity is evaluated at Hubble
crossing. In the time-origin convention used for $H_{\rm app}=2H_1t$, the
expanding branch has $t_*^{\rm(app)}<0$.

At first order, ${n}_s=1-2\widetilde\epsilon_{1*}-\widetilde\epsilon_{2*}$,
where $\widetilde\epsilon_1=-\dot{\widetilde H}/\widetilde H^2$ and
$\widetilde\epsilon_2=d\ln(\widetilde\epsilon_1)/d\widetilde N$.
\begin{equation}
    {n}_s= 1-\frac{3}{t_*^{4} \mathit{H_1}^{2}}+\frac{4 \mathit{H_1} \,t_*^{2}+1}{2 t_*^{4} \mathit{H_1}^{2}}
    \label{eq:ns}
\end{equation}
For illustration, we adopt the pivot scale
\(k_*=0.05\,{\rm Mpc}^{-1}
=1.31\times10^{-58}M_{\rm P}\).
The scalar-amplitude and spectral-index inputs are those specified
in Eqs.~(\ref{eq:nscondition})~and~(\ref{eq:Ascondition}). The tensor bound $r_{0.05}<0.036$ at $95\%$ confidence level is taken
from Ref.~\cite{BICEP:2021xfz}.

For the leading-background benchmark, the observational constraints and
Eqs.~(\ref{eq:kstar}), (\ref{eq:PSinpms}), and~(\ref{eq:ns}) give
\begin{equation}
H_1^{\rm(app)}=-1.2535167949\times10^{-11}M_{\rm P}^2.
\end{equation}
Although the Einstein-frame quantities are written parametrically in terms of
$t$, the latter remains the Jordan-frame cosmic time.  The Einstein-frame
cosmic time is related to it by
\begin{equation}
\widetilde t+\widetilde C_{1}
=
\frac{t}{\sqrt{6}}
\sqrt{1-2H_{1}t^{2}}
+
\frac{1}{2\sqrt{-3H_{1}}}\,
\operatorname{arsinh}\!\left(\sqrt{-2H_{1}}\,t\right),
\qquad H_{1}<0 .
\label{eq:Einstein-time}
\end{equation}
$\widetilde C_1$ is arbitrary, and we set it to zero.  For this
$H_{\rm app}$ benchmark,
\begin{equation}
t_*^{\rm(app)}=-2.1548175399\times10^6M_{\rm P}^{-1},
\qquad
t_f^{\rm(app)}=-2.6284531306\times10^5M_{\rm P}^{-1},
\end{equation}
where $t_f^{\rm(app)}$ follows from
$\widetilde\epsilon_{1,\rm app}=1$.  The scale-factor normalization fixed by
the pivot condition is $a_0=4.6344\times10^{-29}$.  Since $H_1<0$, expansion
requires $t<0$, and physical time runs from more negative to less negative
values.  Thus $t_*^{\rm(app)}<t_f^{\rm(app)}<0$.

The principal corrected background gives, for the Planck 2018 central value
$n_s=0.9649$,
\begin{equation}
(\widetilde N_*,r)_{\rm corr2}
=(55.60241,0.00355214),
\end{equation}
compared with $(55.72453,0.00355212)$ for the converged numerical-background
benchmark and $(55.45745,0.00354225)$ for $H_{\rm app}$.  The errors in
$\widetilde N_*$ relative to the numerical background are therefore about
$0.12$ and $0.27$ e-fold for $H_{\rm corr2}$ and $H_{\rm app}$,
respectively.  The conventional large-$\widetilde N_*$ relations instead
give $(56.98006,0.00369603)$, so their e-fold estimate exceeds the numerical
result by about $1.26$ e-folds.

\[
V(\widetilde\phi)=\frac{M_{\rm P}^2}{8\beta}
\left(1-e^{-\sqrt{2/3}\,\widetilde\phi/M_{\rm P}}\right)^2.
\]

We use
\[
\epsilon_V=\frac{M_{\rm P}^2}{2}\left(\frac{V'}{V}\right)^2,
\quad \eta_V=M_{\rm P}^2\frac{V''}{V},
\quad \xi_V^2=M_{\rm P}^4\frac{V'V'''}{V^2}.
\]
With $x=e^{\sqrt{2/3}\,\widetilde\phi/M_{\rm P}}=F$. Here primes denote derivatives with respect to the Einstein-frame scalar field $\widetilde{\phi}$.
\[
\epsilon_V=\frac{4}{3(x-1)^2},\quad
\eta_V=-\frac{4(x-2)}{3(x-1)^2},\quad
\xi_V^2=\frac{16(x-4)}{9(x-1)^3}.
\]

\begin{figure*}
    \centering
    \includegraphics[width=0.975\linewidth]{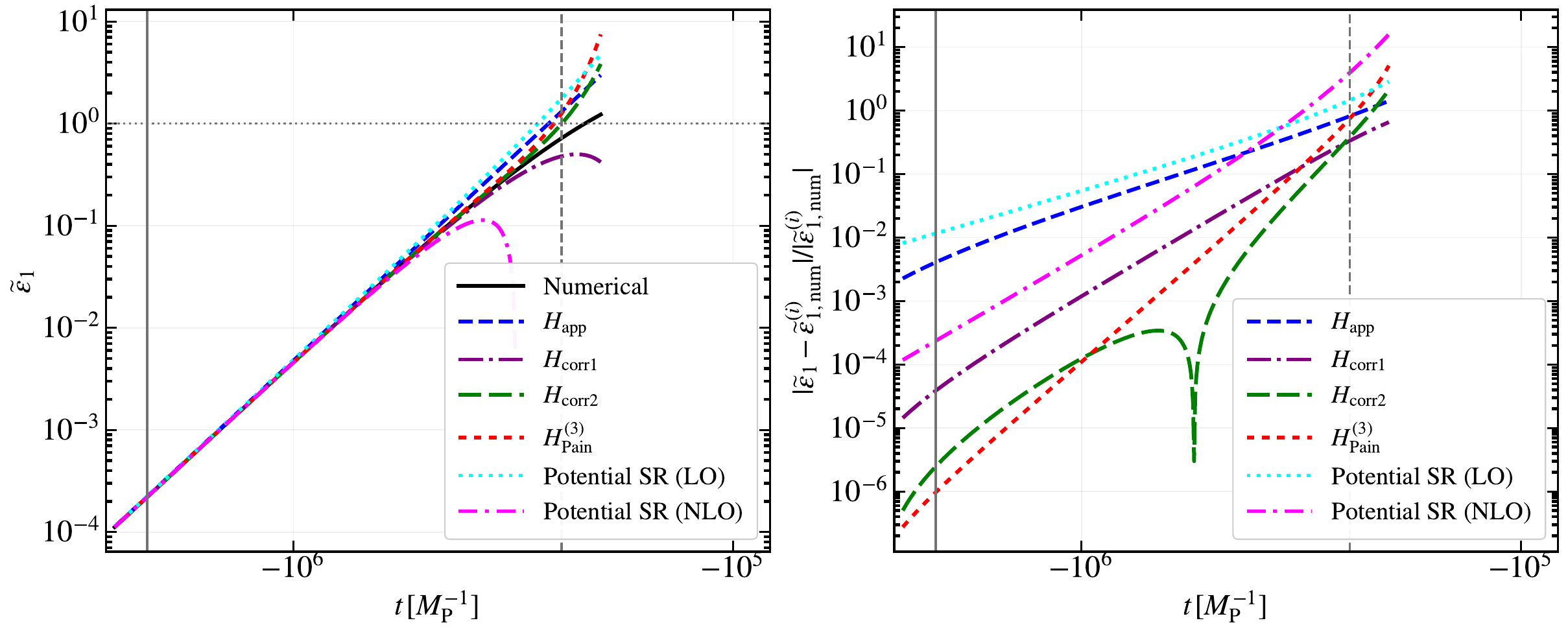}
    \hfill

    \vspace{0.5em}

    \includegraphics[width=0.975\linewidth]{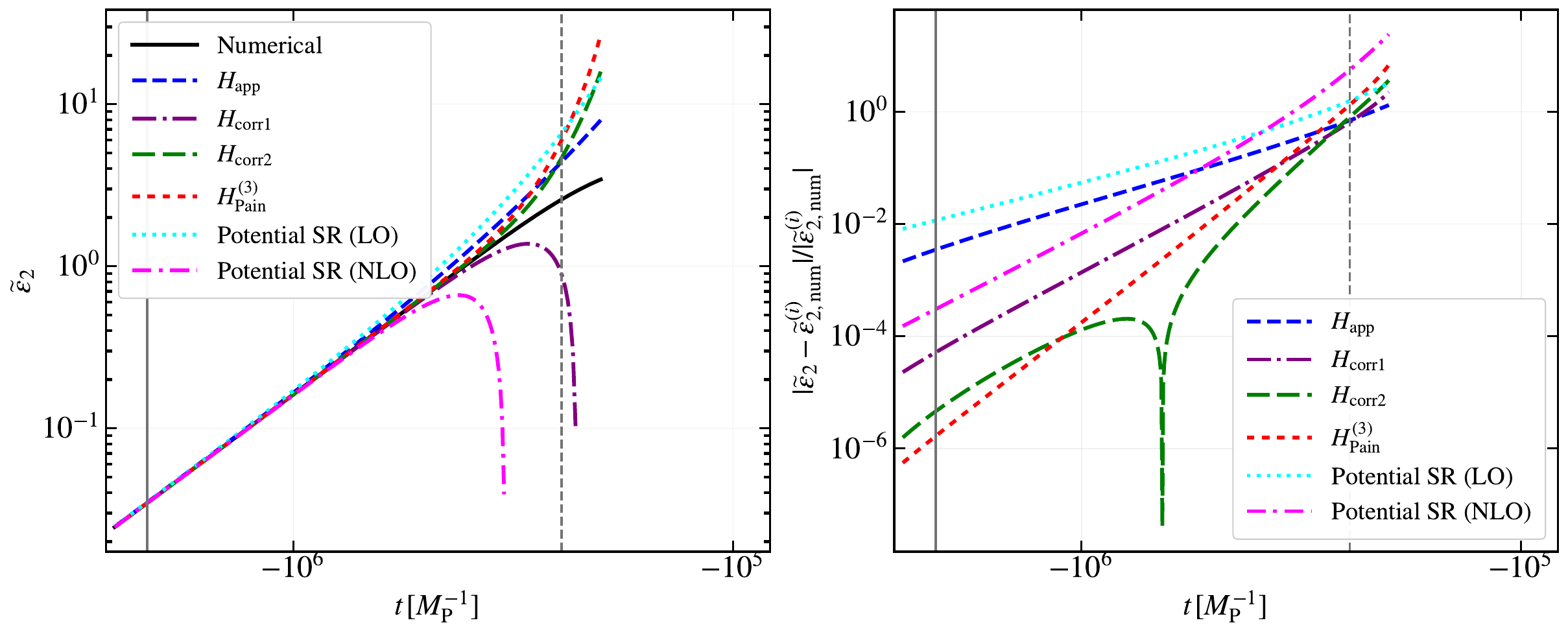}
    \hfill
\caption{
Einstein-frame Hubble-flow parameters and their relative deviations
from converged numerical solutions of the exact pure-Starobinsky
background equation. The upper pair of panels shows
\(\widetilde\epsilon_1\) and its relative error, while the lower pair
shows \(\widetilde\epsilon_2\) and its relative error. In the two
left-hand panels, the black curve labelled ``Numerical'' denotes the
independently calibrated numerical-background benchmark. Each
analytic background uses its own calibrated
\(\bigl(H_1^{(i)},\beta^{(i)},t_*^{(i)}\bigr)\) and translation
parameter. In each right-hand relative-error panel,
\(\widetilde\epsilon_{n,\mathrm{num}}^{(i)}\) denotes the converged
numerical trajectory generated using the same \(\beta^{(i)}\) as
analytic scheme \(i\). Thus, the numerical reference used for an
analytic curve in a relative-error panel is scheme matched and need
not coincide with the single black numerical curve displayed in the
corresponding left-hand panel. The LO- and NLO-PSR quantities are
evaluated along the independently calibrated numerical trajectory.
The grey solid and dashed lines mark
\(t_*^{(\mathrm{corr2})}\) and \(t_f^{(\mathrm{corr2})}\),
respectively. During most of slow roll, NLO PSR is more accurate than
\(H_{\rm app}\), although this ordering is not uniform near the
endpoint. Over the domain in which the plotted NLO expressions remain
positive, \(H_{\rm corr2}\) is more accurate than NLO for both
parameters. The NLO \(\widetilde\epsilon_1\) curve disappears when the
truncated expression becomes nonpositive and never reaches unity.
}
    \label{fig:HvsV}
\end{figure*}

For completeness, the endpoint obstruction in the NLO-PSR row
of Table~\ref{tab:schemes} can be seen analytically. Using the second-order
potential-to-Hubble slow-roll relations of
Ref.~\cite{Liddle:1994dx}, we obtain the following LO and NLO
expressions for the Einstein-frame Hubble-flow parameters
used in Fig.~\ref{fig:HvsV}:
\begin{align}
\widetilde\epsilon_1^{\rm(LO)}
&=\epsilon_V
\label{eq:eps1LO}\\
\widetilde\epsilon_2^{\rm(LO)}
&=4\epsilon_V-2\eta_V,\label{eq:eps2LO}\\ 
\widetilde\epsilon_1^{\rm(NLO)}
&=\epsilon_V-\frac{4}{3}\epsilon_V^2
+\frac{2}{3}\epsilon_V\eta_V,
\label{eq:eps1NLO}\\
\widetilde\epsilon_2^{\rm(NLO)}
&=4\epsilon_V-2\eta_V-8\epsilon_V^2
+\frac{20}{3}\epsilon_V\eta_V
-\frac{2}{3}\eta_V^2-\frac{2}{3}\xi_V^2.\label{eq:eps2NLO} 
\end{align}
On the inflationary branch, Eq.~(\ref{eq:eps1NLO}) reaches its maximum at
$x=(5+2\sqrt{5})/3$, where
$\widetilde\epsilon_1^{\rm(NLO)}=0.11373$.  It subsequently decreases and
becomes negative instead of reaching unity.  Hence this finite NLO
truncation has no self-consistent end-of-inflation solution.
Within LO PSR, Drees and Xu~\cite{Drees:2025ngb} obtain
$\widetilde N_*=54.87588$ by evaluating the e-folding integral without a
large-$\widetilde N_*$ truncation.  Their small-$(1-n_s)$ expansion,
\begin{equation}
\widetilde N_*\simeq
\frac{2}{1-n_s}+\frac{3}{4}\ln(1-n_s)+0.47,
\end{equation}
gives $54.93789$.  This distinction is important: the common
$2/(1-n_s)$ result is a further large-$\widetilde N_*$ truncation, not the
complete LO-PSR calculation.  Our construction is complementary because it
derives the Hubble-flow hierarchy from an analytic background and determines
the endpoint from that same background.


We first examine the dependence of the predictions on the adopted
spectral-index benchmark by considering the Planck 2018 and ACT DR6
combined-data values. For the ACT comparison, only \(n_s\) is changed;
the same adopted scalar-amplitude benchmark is retained. The
conventional large-\(\widetilde N_*\) relations, the Drees--Xu
expansion, \(H_{\rm app}\), and \(H_{\rm corr2}\) are compared in the
following table, Table~\ref{tab:Nstar}.
\FloatBarrier
\begin{table}[!htbp]
\centering

\caption{Comparison of the inflationary observables obtained from the
conventional large-$\widetilde N_*$ approximation, the refined LO-PSR
expansion of Drees and Xu~\cite{Drees:2025ngb}, and the analytic backgrounds
$H_{\mathrm{app}}$ and $H_{\mathrm{corr2}}$. The standard column uses
$\widetilde N_*=2/(1-n_s)$ and $r=12/\widetilde N_*^2$, whereas the
Drees--Xu column uses
$\widetilde N_*=2/(1-n_s)+(3/4)\ln(1-n_s)+0.47$. No value of $r$ is
assigned to the latter relation because it refines the determination of
$\widetilde N_*$ only; substituting this value into the large-$\widetilde
N_*$ relation $r=12/\widetilde N_*^2$ would mix two different
approximations. The analytic results are evaluated self-consistently from
the corresponding Einstein-frame Hubble-flow quantities.}
\label{tab:Nstar}
\resizebox{\textwidth}{!}{
\begin{tabular}{lccccc}
\hline
Data set & Observable & Standard & Drees--Xu expansion
& $H_{\mathrm{app}}$ & $H_{\mathrm{corr2}}$ \\
\hline
Planck 2018~\cite{Planck:2018jri}, $n_s=0.9649$
& $\widetilde N_*$ & $56.98$ & $54.94$ & $55.46$ & $55.60$ \\
& $r$ & $0.003696$ & -- & $0.003542$ & $0.003552$ \\
\hline
ACT DR6 combined-data~\cite{AtacamaCosmologyTelescope:2025blo}, $n_s=0.974$
& $\widetilde N_*$ & $76.92$ & $74.66$ & $75.26$ & $75.43$ \\
& $r$ & $0.002028$ & -- & $0.001965$ & $0.001969$ \\
\hline
\end{tabular}
}
\end{table}

We next perform a detailed scheme-by-scheme comparison for the
adopted Planck benchmark. Table~\ref{tab:schemes} lists the predicted values of
$\widetilde N_*$ and $r$, the scheme-dependent calibration
parameters, and whether each approximation supplies a
self-consistent endpoint.

\begin{table*}[!htbp]
\centering

\caption{
Comparison of the inflationary predictions for the Planck 2018 central
value $n_s=0.9649$. The quantities $H_1$ and $t_*$ are quoted only for
schemes possessing an explicit Jordan-frame background. For entries marked
``No solution'', the number in parentheses denotes the maximum value of
$\widetilde\epsilon_1$, rather than $x_{\rm end}$. For the Painlev\'e
background, the quoted $t_*$ uses the time origin fixed by matching to the
numerical attractor. The final column reports $B_2^{(i)}/M_{\rm P}$ for the corrected backgrounds and $t_0\,M_{\rm P}$ for the Painlev\'e background. 
}
\label{tab:schemes}

\resizebox{\textwidth}{!}{
\begin{tabular}{lcccccc}
\hline
Method
& $H_1/M_{\rm P}^{2}$
& $t_* M_{\rm P}$
& $x_{\rm end}$
& $\widetilde N_*$
& $r$
& \shortstack{$B_2^{(i)}/M_{\rm P}$\\or $t_0M_{\rm P}$}\\
\hline

Conventional large-$\widetilde N_*$ relations
& ---
& ---
& ---
& $56.98006$
& $0.00369603$ 
&---\\

LO PSR, exact e-folding integral
& ---
& ---
& $2.15470$
& $54.87588$
& $0.00351337$ 
&---\\

NLO PSR
& ---
& ---
& No solution $(0.11373)$
& Undefined
& $0.00355339$ 
&---\\

$H_{\rm app}$
& $-1.25351679\times10^{-11}$
& $-2.15481754\times10^{6}$
& $1.82137$
& $55.45745$
& $0.00354225$
&---\\

Painlev\'e, $n=3$
& $-1.25707543\times10^{-11}$
& $-2.14810467\times10^{6}$%
  \textsuperscript{\(\dagger\)}
& $1.91224$
& $55.54709$
& $0.00355211$
& $-2.469516\times10^3$ \\

$H_{\rm corr1}$
& $-1.25715806\times10^{-11}$
& $-2.14797804\times10^{6}$
& No solution $(0.50101)$
& Undefined
& $0.00355234$
&$-6.223923826\times10^{-8}$ \\

$H_{\rm corr2}$ (this work)
& $-1.25708660\times10^{-11}$
& $-2.14808858\times10^{6}$
& $1.84660$
& $55.60241$
& $0.00355214$
& $-6.207073713\times10^{-8}$\\

Numerical background (first spectral order)
& $-1.25707832\times10^{-11}$
& $-2.14810046\times10^{6}$
& $1.65178$
& $55.72453$
& $0.00355212$ 
& ---\\

\hline
\end{tabular}
}

\vspace{1mm}
\parbox{\textwidth}{\footnotesize
\textsuperscript{\(\dagger\)}
For the matched Painlev\'e time origin,
$t_0=-2.46951671\times10^{3}M_{\rm P}^{-1}$.
The shift-invariant horizon-crossing time is
$(t_*-t_0)M_{\rm P}=-2.14563515\times10^{6}$.
The $H_{\rm corr1}$ calibration uses
$B_2^{\rm (corr1)}=-6.22392383\times10^{-8}M_{\rm P}$.
Although its horizon-crossing solution exists,
$\widetilde\epsilon_{1,\rm corr1}$ reaches only
$\max\widetilde\epsilon_{1,\rm corr1}\simeq0.50101$;
therefore it supplies neither $x_{\rm end}$ nor a self-consistent
$\widetilde N_*$.
}
\end{table*}

The conventional large-$\widetilde N_*$ relation gives
$r\simeq0.003696$, whereas the principal corrected background
$H_{\mathrm{corr2}}$ gives $r\simeq0.003552$; the leading benchmark
$H_{\mathrm{app}}$ gives $r\simeq0.003542$. Figure~\ref{fig:HvsV} separates pointwise accuracy from endpoint
completeness. During most of slow roll, NLO PSR is more accurate pointwise than
$H_{\rm app}$, although this ordering is not uniform as the endpoint
is approached. Over the common plotted domain on which the plotted NLO
expressions remain physical, $H_{\rm corr2}$ is more accurate than
NLO for both Hubble-flow parameters. The NLO curve nevertheless cannot determine the end
of inflation because Eq.~(\ref{eq:eps1NLO}) becomes nonpositive after attaining
its small maximum.  By contrast, the endpoint-complete analytic backgrounds $H_{\rm app}$,
$H_{\rm corr2}$, and $H_{\rm Pain}^{(3)}$ each supply their own
solution of $\widetilde\epsilon_1=1$ and hence a self-consistent
integration interval for $\widetilde N_*$. The $H_{\rm corr1}$
truncation is the exception listed in Table \ref{tab:schemes}: its maximum
$\widetilde\epsilon_1$ is below unity, so it has no endpoint.

The numerical curves in Figs.~\ref{fig:Hcomp}~and~\ref{fig:HvsV} are obtained by
integrating the exact pure-Starobinsky Jordan-frame equation as a first-order
system for $(H,\dot H)$; $\ddot H$ is supplied algebraically by
Eq.~(\ref{eq:exactEOM}).  A DOP853 adaptive integrator is used with relative and
absolute tolerances $10^{-13}$ and $10^{-22}$, respectively, and all required
higher derivatives are generated from the differential equation rather than
by numerical differencing.  The resulting trajectory is mapped to the
Einstein frame using Eq.~(\ref{eq:rel2frames}).  Potential slow-roll
quantities in Fig.~\ref{fig:HvsV} are evaluated from the Einstein-frame
potential along this same numerical trajectory.
\paragraph{Tolerance-sensitivity test.}
As an independent convergence check, we repeated the complete numerical
procedure after introducing the dimensionless variables
\begin{equation}
 z=\sqrt{-H_1}\,t,
 \qquad
 g=\frac{H}{\sqrt{-H_1}},
\end{equation}
for which the pure-Starobinsky background equation becomes
\begin{equation}
 g'=p,
 \qquad
 p'=\frac{p^2}{2g}-3gp-6g,
 \qquad p\equiv g',
\end{equation}
where a prime denotes differentiation with respect to $z$. The absolute
tolerances reported in Table~\ref{tab:tolerance_test} therefore apply
to this dimensionless first-order system and should not be identified
directly with the dimensional absolute tolerance quoted above for the
production integration.

For each tolerance pair, the dimensionless pure-Starobinsky Jordan-frame
equation was integrated using the DOP853 method, the resulting trajectory
was mapped to the Einstein frame, and the horizon-crossing calibration
and endpoint determination were repeated independently. Thus, the values
of $H_1$, $t_*$, $t_f$, $\widetilde N_*$, and $r$ were recalculated
rather than being evaluated on a fixed reference trajectory.

Table~\ref{tab:tolerance_test} shows the resulting convergence. The
tightest calculation, with
$(\mathrm{rtol},\mathrm{atol})=(10^{-13},10^{-15})$, is used as the
reference. Over the full tolerance range shown, the largest variations
are
\begin{equation}
 \max\left|\Delta\widetilde N_*\right|
 =4.57\times10^{-5},
 \qquad
 \max\left|\Delta r\right|
 =5.66\times10^{-9}.
 \label{eq:tolerance_variations}
\end{equation}
Already for $\mathrm{rtol}\leq10^{-10}$, the corresponding changes
relative to the tightest calculation are below
$3.5\times10^{-8}$ in $\widetilde N_*$ and
$4.3\times10^{-12}$ in $r$. The numerical-background predictions
quoted in Table~\ref{tab:schemes} are therefore stable under progressive
refinement of the integration tolerances at substantially better
precision than the digits reported in the manuscript.
\begin{table}[!htbp]
\centering
\caption{Tolerance sensitivity of the numerical predictions obtained
using the dimensionless Jordan-frame system. The absolute tolerances
listed here apply to the dimensionless variables, and the differences
are measured relative to the tightest run.}
\label{tab:tolerance_test}
\begin{tabular}{ccccc}
\toprule
$\mathrm{rtol}$ & $\mathrm{atol}$ & Steps &
$\widetilde N_{*}$ & $r$ \\
\midrule
$10^{-9}$  & $10^{-11}$ & 114 & 55.724482091 & 0.003552125745 \\
$10^{-10}$ & $10^{-12}$ & 182 & 55.724527872 & 0.003552120076 \\
$10^{-11}$ & $10^{-13}$ & 266 & 55.724527841 & 0.003552120080 \\
$10^{-12}$ & $10^{-14}$ & 382 & 55.724527838 & 0.003552120080 \\
$10^{-13}$ & $10^{-15}$ & 544 & 55.724527838 & 0.003552120080 \\
\bottomrule
\end{tabular}
\end{table}

\clearpage
\section{Conclusion}
\label{sec:conclusion}

We have constructed the Einstein-frame Hubble-flow hierarchy by mapping
closed-form Jordan-frame Hubble backgrounds, rather than replacing the
hierarchy by derivatives of the potential.  The leading background
$H_{\rm app}$ is an exact solution of a controlled deformation of the
Starobinsky action. Linearization about $H_{\rm app}$ gives the exact error-function
correction in Eq.~(\ref{eq:deltaHexact}). The principal background $H_{\rm corr2}$ is
the corresponding large-$|t|$ asymptotic expansion truncated at order
$t^{-3}$. Its retained alignment constant $B_2^{(\mathrm{corr2})}$ is
fixed by matching the truncated background to the pure-Starobinsky
numerical attractor.

The comparison distinguishes pointwise accuracy from the ability to
determine the endpoint. In the scheme-specific comparison shown in
Fig.~\ref{fig:HvsV}, NLO PSR is more accurate than $H_{\rm app}$ during most of
slow roll, although this ordering is not uniform as the endpoint is
approached. Over the domain in which the truncated NLO expressions
remain physically meaningful, $H_{\rm corr2}$ is more accurate than
NLO PSR for both plotted Hubble-flow parameters. Nevertheless, the truncated NLO expression for 
\(\widetilde{\epsilon}_1\) reaches only a small maximum, subsequently becomes negative, and never satisfies \(\widetilde{\epsilon}_1=1\). It therefore cannot determine \(t_f\) or \(\widetilde N_*\) self-consistently. By contrast, $H_{\rm app}$, $H_{\rm corr2}$, and $H_{\rm Pain}^{(3)}$ each provide a self-consistent endpoint. The $H_{\rm corr1}$ truncation does not, and the residuals of the endpoint-complete analytic backgrounds increase as their endpoints are approached.

For the adopted Planck benchmark, \(H_{\rm corr2}\) reproduces the numerical-background predictions for \(\widetilde N_*\) and \(r\) closely and gives a smaller e-fold discrepancy than \(H_{\rm app}\). The conventional large-\(\widetilde N_*\) relations yield larger values of both quantities. Among the analytic backgrounds compared in Table~\ref{tab:schemes}, \(H_{\rm corr2}\) gives the value of \(\widetilde N_*\) closest to the numerical benchmark, whereas the Painlev\'e result for \(r\) is marginally closer at the displayed precision. The corresponding numerical values and differences are reported in Table \ref{tab:schemes}.

The comparison with LO PSR also demonstrates that the conventional large-\(\widetilde N_*\) formula should not be identified with the complete LO framework. The exact LO-PSR e-folding integral of Drees and Xu and their small-\((1-n_s)\) expansion differ from the conventional large-\(\widetilde N_*\) estimate. The same scheme dependence is found for the adopted ACT DR6 combined-data calibration. Thus, the inferred e-fold number depends on both the background approximation and any additional large-\(\widetilde N_*\) truncation, rather than simply on whether the calculation is described as LO PSR. The corresponding Planck and ACT results are summarized in Table~\ref{tab:Nstar}, while the Planck comparison across all approximation schemes is presented in Table~\ref{tab:schemes}.

All numerical comparisons in this paper are at first spectral order; they
test background and slow-roll approximations rather than higher-order
corrections to the perturbation spectra.  In addition, connecting the inferred
$\widetilde N_*$ to the observed pivot scale requires a post-inflationary
history.  Reheating can itself shift $\widetilde N_*$ by several e-folds
\cite{Dai:2014jja,Cook:2015vqa}, so combining the present background
improvement with a specified reheating scenario is a separate step.

Consequently, the differences in $\widetilde N_*$ discussed here
become physically meaningful only after the reheating prescription is
held fixed or constrained independently. Moreover, the associated
change in $r$ is too small to constitute a presently distinguishable
observational signature by itself. The main significance of the
comparison is therefore theoretical: as measurements of inflationary
observables improve, uncertainties arising from the background
approximation should be controlled before reheating and higher-order
spectral corrections are combined.

The alternative exact solution summarized in
Appendix~\ref{sec:altsol} is included for completeness
and is not used in the inflationary predictions or in the viability
claims of the main analysis. More generally, the matching strategy developed here can be applied to models for which a sufficiently accurate analytic Hubble background is available, provided that the conformal map is regular, the homogeneous zero modes are fixed consistently, and the residual remains controlled over the relevant inflationary interval. 

\appendix

\section{Alternative exact solution in quadratic gravity with a
cosmological constant}
\label{sec:altsol} 
An exact solution of quadratic gravity with a cosmological constant discussed in
Ref.~\cite{Vernov:2019ubo} has the scale factor
$a(t)=a_0\sqrt{-t}\exp(H_1t^2)$ on the real branch $t<0$.  Taking this scale
factor as the reconstruction ansatz, substituting in Eq.~(\ref{eq:Main}) and solving for $\rm F(R(t))$, gives
\begin{equation}
F(R(t))=
\left[
D_1\int\frac{\exp(H_1t^2)}{(-t)^{7/2}}\,dt+D_2
\right]t^2 .
\end{equation}
where $D_1$ and $D_2$ are integration constants.

Selecting the polynomial branch $D_1=0$, the Ricci scalar is
$R=48H_1^2t^2+36H_1$.  With $D_2=-4H_1/3$, integration of
$F=df/dR$ gives
$f(R)=R+\beta R^2+\Lambda_{\rm add}$, where
\begin{align}
\beta=-\frac{1}{72H_1},
\qquad
\Lambda_{\rm add}=-8H_1 .
\end{align}
Here $\Lambda_{\rm add}$ denotes the additive constant in $f(R)$.
The value of $\Lambda_{\rm add}$ is fixed by the
remaining Friedmann constraint. Substituting
$H=1/(2t)+2H_1t$, $F=-4H_1t^2/3$, and
$\beta=-1/(72H_1)$ gives
\[
0
=
3FH^2-\frac{FR-f}{2}+3H\dot F
=
4H_1+\frac{\Lambda_{\rm add}}{2}.
\]
Consequently, $\Lambda_{\rm add}=-8H_1$.
This completes the consistency check beyond the
reconstruction equation~(5).

  If the
action is instead written in the standard form
$f(R)=R+\beta R^2-2\Lambda_{\rm phys}$, then
$\Lambda_{\rm phys}=4H_1$.  This branch is included only to document the
additional exact solution and is not used in the inflationary analysis or in
the viability claims of the main text.
 \section*{Acknowledgements}
The author thank Krishnamohan Parattu and Sandeep K for useful discussions. 
He also thank Manosh T. M, Asha Sebastian and S. Shankaranarayanan for comments on the 
manuscript.

\textbf{Conflict of Interest:} The author declares that there is no conflict 
of interest.

\textbf{Funding:} This research received no specific grant from any funding agency in the 
public, commercial, or not-for-profit sectors.

\textbf{Data Availability Statement:} No external datasets were used in this study. The computational files and numerical data supporting the findings of this study are available from the corresponding author upon reasonable request.

\textbf{Ethics Statement:} Not applicable. This study did not involve human 
participants, animal subjects, or clinical data.
\bibliographystyle{iopart-num}
\bibliography{ref}

@PREAMBLE{
 "\providecommand{\noopsort}[1]{}" 
 # "\providecommand{\singleletter}[1]{#1}%" 
}

@article{khan2023impact,
  title={Impact of complexity factor on anisotropic generalized polytropes in f (R, T) gravity},
  author={Khan, S and Siddiqui, AA and Mardan, SA and Noureen, I},
  journal={The European Physical Journal Plus},
  volume={138},
  number={9},
  pages={1--11},
  year={2023},
  publisher={Springer}
}

@misc{capozziello2011beyond,
  title={Beyond einstein gravity},
  author={Capozziello, Salvatore and Faraoni, Valerio and others},
  year={2011},
  publisher={Springer Netherlands Dordrecht, Holland}
}

@article{Liddle:1994dx,
    author = "Liddle, Andrew R. and Parsons, Paul and Barrow, John D.",
    title = "{Formalizing the slow roll approximation in inflation}",
    eprint = "astro-ph/9408015",
    archivePrefix = "arXiv",
    reportNumber = "SUSSEX-AST-94-8-1",
    doi = "10.1103/PhysRevD.50.7222",
    journal = "Phys. Rev. D",
    volume = "50",
    pages = "7222--7232",
    year = "1994"
}

@article{Mathew:2017lvh,
    author = "Mathew, Jose and Johnson, Joseph P. and Shankaranarayanan, S.",
    title = "{Inflation with $f(R,\phi)$ in Jordan frame}",
    eprint = "1705.07945",
    archivePrefix = "arXiv",
    primaryClass = "gr-qc",
    doi = "10.1007/s10714-018-2410-4",
    journal = "Gen. Rel. Grav.",
    volume = "50",
    number = "7",
    pages = "90",
    year = "2018"
}

@article{Toyama:2024ugg,
    author = "Toyama, Shunsuke and Ketov, Sergei V.",
    title = "{Starobinsky inflation beyond the leading order}",
    eprint = "2407.21349",
    archivePrefix = "arXiv",
    primaryClass = "gr-qc",
    reportNumber = "IPMU24-0033",
    doi = "10.1103/PhysRevD.110.063552",
    journal = "Phys. Rev. D",
    volume = "110",
    number = "6",
    pages = "063552",
    year = "2024"
}

@article{Guth:1980zm,
    author = "Guth, Alan H.",
    editor = "Fang, Li-Zhi and Ruffini, R.",
    title = "{The Inflationary Universe: A Possible Solution to the Horizon and Flatness Problems}",
    reportNumber = "SLAC-PUB-2576",
    doi = "10.1103/PhysRevD.23.347",
    journal = "Phys. Rev. D",
    volume = "23",
    pages = "347--356",
    year = "1981"
}

@article{Linde:1981mu,
    author = "Linde, Andrei D.",
    editor = "Fang, Li-Zhi and Ruffini, R.",
    title = "{A New Inflationary Universe Scenario: A Possible Solution of the Horizon, Flatness, Homogeneity, Isotropy and Primordial Monopole Problems}",
    reportNumber = "LEBEDEV-81-229",
    doi = "10.1016/0370-2693(82)91219-9",
    journal = "Phys. Lett. B",
    volume = "108",
    pages = "389--393",
    year = "1982"
}

@article{Albrecht:1982wi,
    author = "Albrecht, Andreas and Steinhardt, Paul J.",
    editor = "Fang, Li-Zhi and Ruffini, R.",
    title = "{Cosmology for Grand Unified Theories with Radiatively Induced Symmetry Breaking}",
    reportNumber = "UPR-0185T",
    doi = "10.1103/PhysRevLett.48.1220",
    journal = "Phys. Rev. Lett.",
    volume = "48",
    pages = "1220--1223",
    year = "1982"
}

@book{book:15483,
    author = "Liddle, Andrew R. and Lyth, D. H.",
    title = "{Cosmological inflation and large scale structure}",
    doi = "10.1017/CBO9781139175180",
    isbn = "978-0-521-57598-0, 978-0-521-82849-9",
    year = "2000",
    Publisher                = {Cambridge University Press},
}

@article{Mukhanov:1981xt,
    author = "Mukhanov, Viatcheslav F. and Chibisov, G. V.",
    title = "{Quantum Fluctuations and a Nonsingular Universe}",
    journal = "JETP Lett.",
    volume = "33",
    pages = "532--535",
    year = "1981"
}

@article{Hawking:1982cz,
    author = "Hawking, S. W.",
    title = "{The Development of Irregularities in a Single Bubble Inflationary Universe}",
    reportNumber = "Print-83-0015 (CAMBRIDGE)",
    doi = "10.1016/0370-2693(82)90373-2",
    journal = "Phys. Lett. B",
    volume = "115",
    pages = "295",
    year = "1982"
}

@article{Starobinsky:1982ee,
    author = "Starobinsky, Alexei A.",
    title = "{Dynamics of Phase Transition in the New Inflationary Universe Scenario and Generation of Perturbations}",
    doi = "10.1016/0370-2693(82)90541-X",
    journal = "Phys. Lett. B",
    volume = "117",
    pages = "175--178",
    year = "1982"
}

@article{DeFelice:2010aj,
    author = "De Felice, Antonio and Tsujikawa, Shinji",
    title = "{f(R) theories}",
    eprint = "1002.4928",
    archivePrefix = "arXiv",
    primaryClass = "gr-qc",
    doi = "10.12942/lrr-2010-3",
    journal = "Living Rev. Rel.",
    volume = "13",
    pages = "3",
    year = "2010"
}

@article{Starobinsky:1980te,
    author = "Starobinsky, Alexei A.",
    editor = "Khalatnikov, I. M. and Mineev, V. P.",
    title = "{A New Type of Isotropic Cosmological Models Without Singularity}",
    doi = "10.1016/0370-2693(80)90670-X",
    journal = "Phys. Lett. B",
    volume = "91",
    pages = "99--102",
    year = "1980"
}

@article{Whitt:1984pd,
    author = "Whitt, Brian",
    title = "{Fourth Order Gravity as General Relativity Plus Matter}",
    reportNumber = "Print-84-0715 (CAMBRIDGE)",
    doi = "10.1016/0370-2693(84)90332-0",
    journal = "Phys. Lett. B",
    volume = "145",
    pages = "176--178",
    year = "1984"
}

@article{Maeda:1988ab,
    author = "Maeda, Kei-ichi",
    title = "{Towards the Einstein-Hilbert Action via Conformal Transformation}",
    reportNumber = "UTAP-75/88",
    doi = "10.1103/PhysRevD.39.3159",
    journal = "Phys. Rev. D",
    volume = "39",
    pages = "3159",
    year = "1989"
}

@article{Ketov:2022zhp,
    author = "Ketov, Sergei V. and Pozdeeva, Ekaterina O. and Vernov, Sergey Yu.",
    title = "{On the superstring-inspired quantum correction to the Starobinsky model of inflation}",
    eprint = "2211.01546",
    archivePrefix = "arXiv",
    primaryClass = "gr-qc",
    reportNumber = "IPMU22-0053",
    doi = "10.1088/1475-7516/2022/12/032",
    journal = "JCAP",
    volume = "12",
    pages = "032",
    year = "2022"
}

@article{Motohashi:2014tra,
    author = "Motohashi, Hayato",
    title = "{Consistency relation for $R^p$ inflation}",
    eprint = "1411.2972",
    archivePrefix = "arXiv",
    primaryClass = "astro-ph.CO",
    doi = "10.1103/PhysRevD.91.064016",
    journal = "Phys. Rev. D",
    volume = "91",
    pages = "064016",
    year = "2015"
}

@article{Cicoli:2018kdo,
    author = "Cicoli, Michele and De Alwis, Senarath and Maharana, Anshuman and Muia, Francesco and Quevedo, Fernando",
    title = "{De Sitter vs Quintessence in String Theory}",
    eprint = "1808.08967",
    archivePrefix = "arXiv",
    primaryClass = "hep-th",
    doi = "10.1002/prop.201800079",
    journal = "Fortsch. Phys.",
    volume = "67",
    number = "1-2",
    pages = "1800079",
    year = "2019"
}

@book{baumann2022cosmology,
  title={Cosmology},
  author={Baumann, Daniel},
  year={2022},
  publisher={Cambridge University Press}
}

@article{Stewart:1993bc,
    author = "Stewart, Ewan D. and Lyth, David H.",
    title = "{A More accurate analytic calculation of the spectrum of cosmological perturbations produced during inflation}",
    eprint = "gr-qc/9302019",
    archivePrefix = "arXiv",
    reportNumber = "KUNS-1176, LANCS-TH-93-01",
    doi = "10.1016/0370-2693(93)90379-V",
    journal = "Phys. Lett. B",
    volume = "302",
    pages = "171--175",
    year = "1993"
}

@article{Auclair:2022yxs,
    author = "Auclair, Pierre and Ringeval, Christophe",
    title = "{Slow-roll inflation at N3LO}",
    eprint = "2205.12608",
    archivePrefix = "arXiv",
    primaryClass = "astro-ph.CO",
    doi = "10.1103/PhysRevD.106.063512",
    journal = "Phys. Rev. D",
    volume = "106",
    number = "6",
    pages = "063512",
    year = "2022"
}

@article{AtacamaCosmologyTelescope:2025blo,
    author = "Louis, Thibaut and others",
    collaboration = "Atacama Cosmology Telescope",
    title = "{The Atacama Cosmology Telescope: DR6 power spectra, likelihoods and {\ensuremath{\Lambda}}CDM parameters}",
    eprint = "2503.14452",
    archivePrefix = "arXiv",
    primaryClass = "astro-ph.CO",
    reportNumber = "FERMILAB-PUB-25-0071-PPD",
    doi = "10.1088/1475-7516/2025/11/062",
    journal = "JCAP",
    volume = "11",
    pages = "062",
    year = "2025"
}

@article{Drees:2025ngb,
    author = "Drees, Manuel and Xu, Yong",
    title = "{Refined predictions for Starobinsky inflation and post-inflationary constraints in light of ACT}",
    eprint = "2504.20757",
    archivePrefix = "arXiv",
    primaryClass = "astro-ph.CO",
    reportNumber = "MITP-25-033",
    doi = "10.1016/j.physletb.2025.139612",
    journal = "Phys. Lett. B",
    volume = "867",
    pages = "139612",
    year = "2025"
}

@article{Dai:2014jja,
    author = "Dai, Liang and Kamionkowski, Marc and Wang, Junpu",
    title = "{Reheating constraints to inflationary models}",
    eprint = "1404.6704",
    archivePrefix = "arXiv",
    primaryClass = "astro-ph.CO",
    doi = "10.1103/PhysRevLett.113.041302",
    journal = "Phys. Rev. Lett.",
    volume = "113",
    pages = "041302",
    year = "2014"
}

@article{martin2014encyclopaedia,
    author = "Martin, Jerome and Ringeval, Christophe and Vennin, Vincent",
    title = "{Encyclop\ae{}dia Inflationaris}",
    eprint = "1303.3787",
    archivePrefix = "arXiv",
    primaryClass = "astro-ph.CO",
    doi = "10.1016/j.dark.2014.01.003",
    journal = "Phys. Dark Univ.",
    volume = "5-6",
    pages = "75--235",
    year = "2014"
}

@article{martin2014best,
    author = "Martin, J\'er\^ome and Ringeval, Christophe and Trotta, Roberto and Vennin, Vincent",
    title = "{The Best Inflationary Models After Planck}",
    eprint = "1312.3529",
    archivePrefix = "arXiv",
    primaryClass = "astro-ph.CO",
    doi = "10.1088/1475-7516/2014/03/039",
    journal = "JCAP",
    volume = "03",
    pages = "039",
    year = "2014"
}

@article{Cook:2015vqa,
    author = "Cook, Jessica L. and Dimastrogiovanni, Emanuela and Easson, Damien A. and Krauss, Lawrence M.",
    title = "{Reheating predictions in single field inflation}",
    eprint = "1502.04673",
    archivePrefix = "arXiv",
    primaryClass = "astro-ph.CO",
    doi = "10.1088/1475-7516/2015/04/047",
    journal = "JCAP",
    volume = "04",
    pages = "047",
    year = "2015"
}

@article{Hwang:1996xh,
    author = "Hwang, Jai-chan and Noh, Hyerim",
    title = "{Cosmological perturbations in generalized gravity theories}",
    reportNumber = "PRINT-96-116 (KYUNGPOOK)",
    doi = "10.1103/PhysRevD.54.1460",
    journal = "Phys. Rev. D",
    volume = "54",
    pages = "1460--1473",
    year = "1996"
}

@article{noureen2022development,
  title={Development of local density perturbation scheme in f (R) gravity to identify cracking points},
  author={Noureen, I and Arshad, N and Mardan, SA},
  journal={The European Physical Journal C},
  volume={82},
  number={7},
  pages={621},
  year={2022},
  publisher={Springer}
}

@article{mardan2020new,
  title={New models of charged anisotropic polytropes with radiation density},
  author={Mardan, SA and Siddiqui, AA and Noureen, I and Jamil, RN},
  journal={The European Physical Journal Plus},
  volume={135},
  number={1},
  pages={1--18},
  year={2020},
  publisher={Springer}
}

@article{Bianchi:2024qyp,
    author = "Bianchi, Eugenio and Gamonal, Mauricio",
    title = "{Primordial power spectrum at N3LO in effective theories of inflation}",
    eprint = "2405.03157",
    archivePrefix = "arXiv",
    primaryClass = "gr-qc",
    doi = "10.1103/PhysRevD.110.104032",
    journal = "Phys. Rev. D",
    volume = "110",
    number = "10",
    pages = "104032",
    year = "2024"
}

@article{BICEP:2021xfz,
    author = "Ade, P. A. R. and others",
    collaboration = "BICEP, Keck",
    title = "{Improved Constraints on Primordial Gravitational Waves using Planck, WMAP, and BICEP/Keck Observations through the 2018 Observing Season}",
    eprint = "2110.00483",
    archivePrefix = "arXiv",
    primaryClass = "astro-ph.CO",
    doi = "10.1103/PhysRevLett.127.151301",
    journal = "Phys. Rev. Lett.",
    volume = "127",
    number = "15",
    pages = "151301",
    year = "2021"
}

@article{Dimitrijevic:2020dzo,
    author = "Dimitrijevic, Ivan and Dragovich, Branko and Koshelev, Alexey S. and Rakic, Zoran and Stankovic, Jelena",
    title = "{Some Cosmological Solutions of a New Nonlocal Gravity Model}",
    eprint = "2006.16041",
    archivePrefix = "arXiv",
    primaryClass = "gr-qc",
    doi = "10.3390/sym12060917",
    journal = "Symmetry",
    volume = "12",
    number = "6",
    pages = "917",
    year = "2020"
}

@article{Planck:2018jri,
    author = "Akrami, Y. and others",
    collaboration = "Planck",
    title = "{Planck 2018 results. X. Constraints on inflation}",
    eprint = "1807.06211",
    archivePrefix = "arXiv",
    primaryClass = "astro-ph.CO",
    doi = "10.1051/0004-6361/201833887",
    journal = "Astron. Astrophys.",
    volume = "641",
    pages = "A10",
    year = "2020"
}

@article{Vernov:2019ubo,
    author = "Vernov, S. Yu. and Ivanov, V. R. and Pozdeeva, E. O.",
    title = "{Superpotential Method for $F(R)$ Cosmological Models}",
    eprint = "1912.07049",
    archivePrefix = "arXiv",
    primaryClass = "gr-qc",
    doi = "10.1134/S1063779620040735",
    journal = "Phys. Part. Nucl.",
    volume = "51",
    number = "4",
    pages = "744--749",
    year = "2020"
}

@article{Mathew:2016anx,
    author = "Mathew, Jose and Shankaranarayanan, S.",
    title = "{Low scale Higgs inflation with Gauss{\textendash}Bonnet coupling}",
    eprint = "1602.00411",
    archivePrefix = "arXiv",
    primaryClass = "astro-ph.CO",
    doi = "10.1016/j.astropartphys.2016.07.004",
    journal = "Astropart. Phys.",
    volume = "84",
    pages = "1--7",
    year = "2016"
}
\end{document}